\documentclass[12pt,preprint]{aastex}

\newcommand{\sneia}{SNe~Ia}

\slugcomment{Accepted for Publication in The Astrophysical Journal (September 20, 2008 issue)}

\shorttitle{Age and Metallicity Paper}
\shortauthors{Gallagher et al.}

\begin{document}

\title{Supernovae in Early-Type Galaxies: Directly Connecting Age and Metallicity with Type Ia Luminosity}

\author{Joseph S. Gallagher and Peter M. Garnavich}
\affil{Department of Physics, University of Notre Dame, 225 Nieuwland 
Science Hall, Notre Dame, IN 46556-5670}
\email{jgallag3@nd.edu}
\author{Nelson Caldwell and Robert P. Kirshner}
\affil{Harvard-Smithsonian Center for Astrophysics, 60 Garden Street, 
Cambridge, MA 02138}
\author{Saurabh W. Jha}
\affil{Dept. of Physics and Astronomy, Rutgers, the State
University of New Jersey, 136 Frelinghuysen Road, Piscataway, NJ 08854}
\author{Weidong Li, Mohan Ganeshalingam, and Alexei V. Filippenko}
\affil{Department of Astronomy, University of California, Berkeley, CA 
94720-3411}

\begin{abstract}
We have obtained optical spectra of 29 early-type (E/S0) galaxies that hosted type Ia 
supernovae (SNe~Ia).  We have measured absorption-line strengths and compared them to a 
grid of models to extract the relations between the supernova properties and the luminosity-weighted 
age/composition of the host galaxies.  Such a direct measurement is a marked improvement 
over existing analyses which tend to rely on general correlations between the properties of 
stellar populations and morphology. Our galaxy sample ranges over a factor of ten in iron 
abundance and shows both old and young dominant population ages. The same analysis was 
applied to a large number of early-type field galaxies selected from the SDSS spectroscopic 
survey. We find no difference in the age and abundance distributions between the field galaxies 
and the SN~Ia host galaxies. We do find a strong correlation suggesting that SNe~Ia in galaxies whose 
populations have a characteristic age greater than 5 Gyr are $\sim$ 1 mag fainter at $V_{max}$ 
than those found in galaxies with younger populations.  However, the data cannot discriminate 
between a smooth relation connecting age and supernova luminosity or two populations of SN~Ia 
progenitors.  

We find that SN~Ia distance residuals in the Hubble diagram are correlated with host-galaxy 
metal abundance, consistent with the predictions of \citet{tbt03}.  The data show that high iron 
abundance galaxies host less-luminous supernovae.  We thus conclude that the time since 
progenitor formation primarily determines the radioactive Ni production while progenitor 
metal abundance has a weaker influence on peak luminosity, but one not fully corrected by 
light-curve shape and color fitters.  This result, particularly the secondary dependence on 
metallicity, has significant implications for the determination of the equation-of-state parameter, 
$w = P/(\rho c^2)$, and could impact planning for future dark-energy missions such as JDEM.  
 
Assuming no selection effects in discovering SNe~Ia in local early-type galaxies, we find a higher 
specific SN~Ia rate in E/S0 galaxies with ages below 3~Gyr than in older hosts. The higher rate 
and brighter luminosities seen in the youngest E/S0 hosts may be a result of recent star formation and 
represents a tail of the ``prompt" SN~Ia progenitors.
\end{abstract}


\section{Introduction}

Type Ia supernovae (\sneia) have proven to be effective standardizable candles and their high peak 
luminosity has made them excellent cosmological probes.  An empirically determined relation between 
light-curve properties and peak absolute magnitude \citep{P93,H96b,RPK96,gold01,guy07,jrk07,c08} has 
yielded improved estimates of the global Hubble constant \citep{Jha99,Fre01,R05,Sand06}.  Although the 
earliest attempts to measure the matter density of the universe preceded these light-curve shape methods 
\citep{nn89}, and the initial efforts by the Supernova Cosmology Project (SCP) led to incorrect conclusions 
about the cosmic matter density \citep{Perl97}, early work by both the SCP and the High-Z Supernova 
Search Team \citep{Schmidt98} found evidence for low matter density \citep{Perl98,G98a}.  

Subsequent publications
by \citet{R98} and \citet{Perl99} came to the surprising conclusion that the universe is expanding at an 
accelerating rate, driven by a mysterious dark energy; see \citet{k02}, \citet{fili05b}, or \citet{fth08} for
reviews. This surprising result has been confirmed by more-recent supernova observations \citep{Knop03,
tonry03,barris04,R04} and by complementary measurements of the cosmic microwave background (CMB) 
anisotropies \citep[e.g.,][]{Bennett03} and large-scale structure \citep[e.g.,][]{peacock01}.  Focus has now 
shifted from demonstrating the existence of dark energy to constraining its properties \citep{G98b,astier06,
R07,Wood07,mik07,e07,frieman07} and sharpening the tools available for its study \citep[e.g.,][]{kevin07,
wv08}.

Analyses of their spectra and light curves have led to the consensus that SNe~Ia are well characterized 
by a thermonuclear disruption of a carbon-oxygen white dwarf (WD).  The transition from a dynamically stable 
WD to an explosion capable of outshining an entire galaxy is precipitated by the WD mass approaching 
the Chandrasekhar mass limit through matter accretion from a close binary companion.  
The nature of this accretion process is still uncertain, though two scenarios have become the focus of 
SN Ia progenitor modeling.  The first is the single-degenerate (SD) model in which a WD accretes matter 
from a non-degenerate companion \citep{wi73,n82}, while the second is the double-degenerate (DD) model 
involving the merger to two WDs \citep{it84,w84}.  Though promising, our understanding of these models 
remains incomplete with questions persisting about the nature of the binary 
companion (main sequence, asymptotic giant branch, etc.) and accretion flow in the SD scenario, the apparent likelihood that a WD merger 
will result in an accretion-induced collapse rather than a SN~Ia, and the potential for 
multiple stellar-population-dependent channels leading to SN~Ia explosions.

Differentiating among dark-energy models requires mapping the expansion history of the
universe with high precision and a small systematic error. Applying SNe~Ia to this problem
is complicated by the uncertain nature of the explosion physics and progenitor stars \citep[e.g.,][]{how06,hick07}.
Even with the most basic assumption, that a SN~Ia results from the thermonuclear disruption of a near-Chandrasekhar
mass carbon-oxygen WD, there remain potential dangers in blindly applying SNe~Ia to the 
dark-energy problem. Theoretical models \citep{hof98,um99,tbt03} suggest that population age and metallicity may have an 
effect on the WD composition and influence the SN~Ia peak luminosity. Since the average age and 
metal content of stars have evolved over cosmic time, the ensemble character of SN~Ia explosions may be a 
function of lookback time. But the local universe contains a range of stellar ages and metallicities, so an 
empirical calibration of these effects may be possible.

\citet{H96b} were the first to note a correlation between SN~Ia decline rate and host-galaxy 
morphology.  The results showed that intrinsically faint events occur in early-type (E/S0) 
galaxies, while luminous events are often hosted by late-type galaxies. This correlation with morphology 
has since been confirmed in larger SN~Ia samples \citep[e.g.,][]{Gallagher05}, but the cause of the trend 
remains unknown. The existence of a correlation is important because it demonstrates that the SN~Ia 
environment imprints itself on some aspect of the progenitor. On average, spiral and elliptical galaxies 
differ in population age, star-formation rate (SFR), and metal content, with significant overlap in these 
properties across the morphological types. \citet{hamuy00} measured the metallicity of a small sample of 
star-forming hosts to test if age or metal content was the factor influencing SN~Ia luminosity, but the results 
were ambiguous.

\citet{Gallagher05} expanded the number of observed hosts to test the impact of stellar environment on the 
photometric properties of nearby SNe~Ia. Potential dependencies of SN decline rate on host-galaxy absolute 
magnitude and star-formation history were investigated in a sample of galaxies with a broad range of properties
and across the full Hubble sequence.  Metallicity was measured from emission lines, limiting this aspect of the 
study to star-forming hosts. No significant correlation between metallicity and decline rate was detected, though a 
tenuous correlation was found between the metal content and Hubble residuals.  Furthermore, only galaxies 
without significant star formation were found to host faint, fast-declining SNe~Ia. The division by star-formation 
history was more stark than by morphology alone, suggesting that the observed luminosity spread in SNe~Ia results from 
the influence of population age on the WD progenitors. The luminosity/star-formation relation has also been 
seen in a high-redshift sample of SNe~Ia studied by the Supernova Legacy Survey (SNLS) \citep{s06}.

The specific SN~Ia rate is strongly dependent on the amount of current star formation \citep{m05,s06}. 
Actively star-forming galaxies host approximately 10 times as many SNe~Ia per unit mass than do their low star-forming counterparts. \citet{sb05} find that the SN~Ia rate is well described by a combination of two 
distinct populations of SNe~Ia: a ``prompt" (shorter delay time) SN~Ia component dependent upon the recent star
formation in the galaxy and an ``delayed" or ``tardy'' (longer delay time) component dependent on the number of 
low-mass stars.  Assuming the light-curve shape correction methods are perfect, then this will not have an effect 
on the recent cosmological results.  However, if the corrected photometric properties of these two SN populations 
differ (e.g., a variation due to age or metallicity), then it could have implications for derived cosmological parameters because
the short delay time SNe are expected to dominate at high redshift whereas the long delay time events are expected to 
dominate at low redshift.

Here we present our study on the luminosity-weighted ages and metal abundances of early-type host galaxies.  
Early-type galaxies, unlike star-forming spirals, have been shown to host both faint and bright 
SNe~Ia, thereby allowing us to sample the full diversity of SNe~Ia.  Consequently, we focus our study 
on absorption-line spectra of host galaxies and utilize single age, single metallicity stellar population models 
to characterize the stellar populations \citep{v96,v99a,bva01,Vazdekis03,v08}.  The impact of both age 
and metallicity on SN peak magnitude is studied and the sample age and metallicity distributions will be 
compared to absorption-line spectra of field galaxies obtained from the Sloan Digital Sky Survey (SDSS).  
Furthermore, we will infer from our data a measurement of the relative SN~Ia rate as a function of 
galaxy age.  Finally, we will search for second-order dependencies at the level of the SN~Ia intrinsic 
scatter by comparing host-galaxy ages and metallicities to the SN residuals from the Hubble diagram. 

We detail our observational techniques and data-reduction pipeline in \S 2 of this paper.  In \S 3 we report the 
methods for performing emission-line corrections, and for obtaining age and metallicity estimates of our host 
sample. Our results are presented in \S 4, and the summary and conclusions are in \S 5.   

\section{Observations and Data Reduction}

\subsection{Observations}
The objective of this study was to obtain absorption-line spectra of SN~Ia host galaxies with the intent of 
estimating their ages and metallicities through a comparison with the simple stellar population (SSP) 
models of Alexandre Vazdekis \citep{v08,Sanchez-Blazquez06}.  To this end, we obtained spectra of a 
sample of SN Ia host galaxies with the 1.5~m Tillinghast telescope located at the F. L. Whipple 
Observatory.  We employed the FAST spectrograph \citep{Fab98} fitted with the 300 line mm$^{-1}$ 
reflection grating and a $3\arcsec \times 3\arcmin$ slit.  The setup gave a resolution of $\sim 7$~\AA\ 
full width at half maximum intensity (FWHM).

The spectra were obtained over the course of  6 nights during April and July of 2005. Our set was compiled 
from the host galaxies of SNe~Ia from the samples of \citet{Phil99}, \citet{Jha02}, and \citet{gan08}.  Suitable 
targets were those classified as having early type (S0--E), as well as spiral galaxies containing strong absorption lines 
in their spectra (and minimal emission lines) as determined by \citet{Gallagher05}.  In addition to the host galaxies, 
we also obtained spectra for a set of comparison elliptical galaxies having age and metallicities estimates 
determined by \citet{trager00}.

The slit was aligned along the galaxy's major axis with position angles (PAs) determined via the Digital Sky 
Survey (DSS) plates.  In a few cases the PA was altered to prevent a nearby star from falling on the slit.   
Exposure times were varied depending on the galaxy brightness, with a target signal-to-noise ratio (S/N) of 
10--20.  For each target we obtained three spectra that we subsequently combined to improve the S/N and 
remove cosmic rays.  Bias, dark, and dome-flat exposures were taken at the beginning and the end of each 
night, and a comparison spectrum of a He-Ne-Ar lamp was acquired before each target for wavelength 
calibration.  Finally, flux standard star spectra were taken each night with the slit oriented along the parallactic 
angle \citep{fili82}.  

Our full sample of galaxies can be seen in Table 1.  Columns (1) and (2) give the host-galaxy name and hosted 
SN, respectively.  Column (3) provides the peak magnitudes of the hosted SN~Ia, determined from the SN
light-curve data and employing MLCS2k2 \citep{jrk07}.  Column (4) gives the morphological classification 
of each galaxy, while Column (5)  reports the PA for the each observation.  Column (6) shows the width of 
the extracted aperture encompassing all the light out to approximately the effective radius, $R_{e}$\footnote{The 
radius within which half of the total galactic light is contained.}.  Finally, Column (7) gives the source of 
the SN light-curve: (1) the CfA sample \citep{jrk07}, (2) the Katzman Automatic Imaging Telescope (KAIT) 
\citep{gan08,fili01,fili05a}, and (3) SN 2005bl \citep{tau08,g05a}.

\subsection{Data Reduction}
The majority of data reduction for this study was performed using standard techniques 
within IRAF\footnote{IRAF is the Image Reduction and Analysis Facility, a general purpose
 software system for the reduction and analysis of astronomical data. IRAF is written and 
supported by the IRAF programming group at the National Optical Astronomy Observatories 
(NOAO) in Tucson, AZ. NOAO is operated by the  Association of Universities for 
Research in Astronomy (AURA), Inc. under cooperative agreement with the National Science 
Foundation (NSF).}.  The data were bias and dark-subtracted, and flat-fielded to remove the pixel-to-pixel 
variations in the CCD; also, the bad rows and columns of the CCD were interpolated across 
using the routine FIXPIX.  Following preprocessing, the task was to combine the set of three 
spectral images that we had for each target.  The first step was the removal of cosmic rays.  In the 
interest of preserving every image in a set, thereby increasing our statistics, we removed each 
cosmic ray individually using the IMEDIT routine in IRAF.  Cosmic rays that fell on the galaxy 
nucleus itself were left alone until after the aperture extraction. 

One-dimensional aperture extraction was performed using the APALL routine with extraction 
radii equal to the effective radius ($R_{e}$) of each galaxy.  The effective radii were determined 
through an IDL code written to integrate over the average profile of 400 centrally located columns 
along the spatial axis of each image.  Each spectrum was then individually checked for cosmic rays 
landing on one of our relevant absorption-line indices or the surrounding continua.  In the event that 
an interfering cosmic ray was found, steps were taken to either remove it interactively within SPLOT, 
or to remove the image altogether if the cosmic ray was too disruptive.  The average of each set of 
images was generated using IM-COMBINE, and the resultant spectra were wavelength and flux
calibrated.  Finally, the spectra were corrected for Galactic extinction using the IRAF routine 
DEREDDEN \citep{ccm89}.
 
\section{Disentangling Age and Metallicity}
\subsection{Background}
A well known and much reviled puzzle in the study of stellar populations is the observed similarity 
between the effects of age and metallicity on the integrated light of stellar populations \citep[e.g.,][]{Worthey94}.  
Astronomers attempt to unravel this age/metallicity degeneracy through the development of stellar population 
models.  The simplified goal of stellar population synthesis modeling is to find a combination of stars for 
which the integrated spectrum of the stars matches the observed spectrum of the population under study.

Early empirical techniques such as quadratic programming devised by \citet{Faber72} have given way 
to evolutionary population synthesis models which have been improving since the 1970s \citep{TinandGunn76,
Tinsley80,GST81,Worthey94,Buzzoni95,BCT96,Maraston98,bc03,FR97,VL05,Vazdekis03,Delgado05}.  
Whereas empirical models are constructed from a combination of stellar spectra or the spectra of stellar 
clusters, evolutionary synthesis models utilize theoretical stellar evolutionary isochrones as the primary 
constituent. A single isochrone on the Hertzsprung-Russell (HR) diagram gives the locus of luminosities 
and temperatures at a single moment in time for stars of all masses.  Together with an assumed initial mass
function (IMF), an isochrone can be built to accurately model a single-age stellar population.  The final 
model is obtained by converting isochrone parameters to observed spectra and finally integrating along 
the isochrone.   

The populations that are modeled using the above techniques are known as simple stellar populations, 
or SSPs.  SSPs are, by definition, a population of stars created during a single burst event (i.e., all 
the stars have the same age) and possessing a single global metallicity.  However, early-type 
galaxies are often comprised of multiple-component stellar populations.  Galaxy ages derived 
through a comparison with these single-component SSPs are luminosity weighted, and 
are typically sensitive to the youngest component in a real early-type galaxy.  Consequently, it is 
appropriate to view the ages derived in studies such as ours as lower-limit estimates 
\citep{renzini06}.

The determination of age and metallicity via galactic spectra is fairly simple, in principle.  We can exploit
predictions made by population synthesis models to relate physical properties, such as age and metallicity, 
to observables, such as the absorption-line strengths for the strongest atomic and molecular absorption 
features in the optical range.  Measuring the relative strengths of these absorption features, we can estimate 
the age and metallicity of the stellar population that created the spectrum.  

\subsection{Model Broadening}

For the age and metallicity estimates of our galaxies, we employ the single age, single metallicity 
stellar population models described by \citet{v96}, \citet{v99a}, \citet{bva01}, \citet{Vazdekis03}, and \citet{v08}.  The models make 
use of the empirical stellar spectral library MILES \citep{Sanchez-Blazquez06,cen07}.  MILES is a marked 
improvement over previous libraries; it has 985 stars, a spectral resolution of 2.3 \AA\ (FWHM), and a 
wavelength range of 3525--7500 \AA.  

One advantage of empirical libraries over theoretical 
stellar libraries is that they are based on actual stellar spectra and are not dependent on the potential inaccuracies 
and underlying assumptions inherent in any theoretical model.  However, in order for an empirical stellar library 
to sufficiently cover the parameter space in temperature, gravity, abundance, and [$\alpha$/Fe], the stellar 
observations must be of the highest quality.  Consequently, empirical stellar libraries have traditionally been
restricted to the nearest stars, thereby introducing a bias of atmospheric parameters toward those which are 
seen in the solar neighborhood.  The MILES library acquired atmospheric parameters from the literature and 
calibrated the set through a subset of reference field stars from \citet{skc98}; for full details see \citet{cen07}.
In this way, MILES has optimized its stellar atmosphere coverage and improved upon previous libraries.  
Nevertheless, because early-type galaxies are generally $\alpha$ enhanced and metal rich, the potential bias 
toward solar abundance ratios should be kept in mind.       
  
The stellar population synthesis models derived from MILES consist of single age, 
single metallicity spectral energy distributions (SEDs) that adopt a standard Salpeter IMF \citep{sal55}.  
The SEDs possess a FWHM resolution of 2.3~\AA\ and a spectral range similar to our data at 3540--7410~\AA.  
There are 276 models with ages ranging from 0.10 to 17.78 Gyr and metallicities ranging from [M/H] = $-1.68$ 
to 0.20.

In the past, the most widely used method of estimating stellar population age and metallicity has been 
through a comparison of absorption-line indices with those of the Lick/IDS System \citep{Worthey94}.  
Since the stars of the Lick/IDS stellar library are not flux calibrated, data must be converted to the 
instrumental response curve of the original data set \citep{wo97}.  With the advent of improved, 
flux-calibrated stellar libraries, it became possible to generate complete SEDs of a single stellar population \citep{j97,
Sanchez-Blazquez06,va99}, thereby eliminating the need for such a correction.  Furthermore, in order to 
use the Lick population models, it is necessary to transform the observational data to match the resolution of the
Lick spectrograph ($\sim 8$--10~\AA\ FWHM).  For our analysis, the model SEDs are at a higher resolution
than the data; consequently, the models were required to be broadened to the resolution of our data rather 
than the other way around.

Thus, the first step in the analysis was to characterize the wavelength-dependent resolution of our spectra.  
This was accomplished by measuring the line widths of emission lines in the He-Ne-Ar lamp across our 
full spectral range.  A polynomial fit to the resulting scatter plot provided a wavelength-dependent resolution 
function for the FAST spectrograph. We derived a wavelength-dependent broadening term with a square equal to 
the difference between the square of the host-galaxy resolution and that of the resolution of the Vazdekis 
models, given as 2.3~\AA.  

The next broadening term we considered was from the velocity dispersion of the host galaxies.  To 
determine the velocity dispersion for each galaxy, we obtained spectra of velocity standard stars HD12623 
and HD52071; these were taken within a few months of our observing 
runs on the Tillinghast telescope using the FAST spectrograph.  In total, we obtained 6 spectra of 
HD12623 and 14 spectra of HD52071.  Cross-correlation analysis was then performed between these 
templates and the host-galaxy spectra using the IRAF routine FXCOR.  A velocity dispersion and redshift 
estimate were obtained using each template-galaxy combination, with the final result for a given host 
being the average of the measurements made with each of the twenty templates.  

Our two broadening 
terms were summed in quadrature, yielding the final width of our Gaussian kernel used in the convolution 
with the host galaxy.  Each host galaxy had a unique final broadening term and the Vazdekis model SEDs 
were broadened independently according to this term.  In this way, we created separate age/metallicity grids 
for each host galaxy.  The velocity dispersions and heliocentric velocities resulting from this analysis 
are shown in Table~\ref{tbl-2}.  These heliocentric velocities were converted to CMB rest frame for the 
subsequent analysis using the NED velocity calculator.

The age dependent absorption index chosen for this study was H$\beta$. H$\beta$ is known to be  
metallicity insensitive and older stellar populations are known to have relatively strong H$\beta$ absorption, 
eliminating the need for high S/N \citep{card03,Cald03}.  Although our sample was comprised mostly of 
older stellar populations, there was a chance for H$\beta$ emission contamination.  Consequently, 
our data were emission-line corrected with the procedure discussed in the next section.  We chose two 
age-insensitive Fe absorption lines for our abundance analysis, Fe$\lambda$5270 and Fe$\lambda$4383.  
Plotting both of these iron lines against H$\beta$ produces a strongly orthogonal grid well suited for 
untangling the age-metallicity degeneracy.  Our chosen Lick/IDS indices are summarized in 
Table~\ref{tbl-3}.

\subsection{Emission-Line Correction}

The presence of H$\beta$ emission superimposed on H$\beta$ absorption poses a significant problem 
for our index measurements.  The contamination of even weak H$\beta$ emission lines in our galaxy spectra 
could significantly decrease the strengths of our measured H$\beta$ absorption indices, thus systematically
biasing our age distribution toward greater ages.  Consequently, an emission correction was performed on 
our host-galaxy spectra.  The basic procedure is outlined by \citet{Cald03}. The method is based on two 
main assertions.  First, although the underlying absorption-line spectrum at H$\beta$ is highly sensitive to age, 
the sensitivity to both age and metallicity is negligible at H$\alpha$.  Second, the observed Hn/Fe 
index\footnote{Hn/Fe $\equiv$ $<$H$\delta$/Fe$\lambda$4045 +H$\gamma$/Fe$\lambda$4325 + 
H8/Fe$\lambda$3859$>$.} represents an effective indicator of the true underlying absorption-line spectrum, 
implying that two galaxies with similar Hn/Fe indices will likewise possess similar H$\alpha$ absorption 
strengths. 

Conservatively, we assumed that each galaxy in our sample was contaminated with Balmer emission from H~II 
regions not included in the Vazdekis models.  Therefore, we predicted the underlying H$\alpha$ absorption 
spectrum for a given contaminated galaxy spectrum by matching that galaxy's Hn/Fe index with a Vazdekis 
model galaxy displaying a similar Hn/Fe index.  The matching Vazdekis model SED was then treated as 
our model absorption spectrum.  Next, we smoothed this spectrum to the appropriate resolution and velocity 
dispersion, normalized it to zero, and subtracted it from the host-galaxy spectrum.  This left the H$\alpha$ 
emission free of any contaminating absorption.  The emission-line flux was measured and converted into H$\beta$ 
emission flux assuming a Case B Balmer decrement \citep{Ost89}. 

We used a flux-calibrated spectrum of the Orion Nebula taken with the same instrumental setup to model the 
emission spectrum intended to be subtracted off the contaminated galaxy.  The Orion spectrum was smoothed
to the appropriate velocity dispersion and scaled so that the H$\beta$ flux matched the calculated value.  The 
continuum was removed from the modified Orion spectrum, and the result was subtracted from the contaminated 
host-galaxy spectrum.

\section{Results}

\subsection{Diagnostic Grids}

In order to measure the H$\beta$, Fe$\lambda$5270, and Fe$\lambda$4383 indices, we employed the 
FORTRAN77 code LECTOR provided by Alexandre Vazdekis.  The code measures the indices 
of a one-dimensional input spectrum and calculates an error estimate based on photon statistics 
\citep{card98, va99, cen01}.   With the line strengths we generated our index-index diagrams, in 
which each vertex represents a combination of line-index measurements for a single model of given 
age and metallicity.  

\begin{figure}[h]
\begin{center}
\includegraphics[angle=0,scale=0.6]{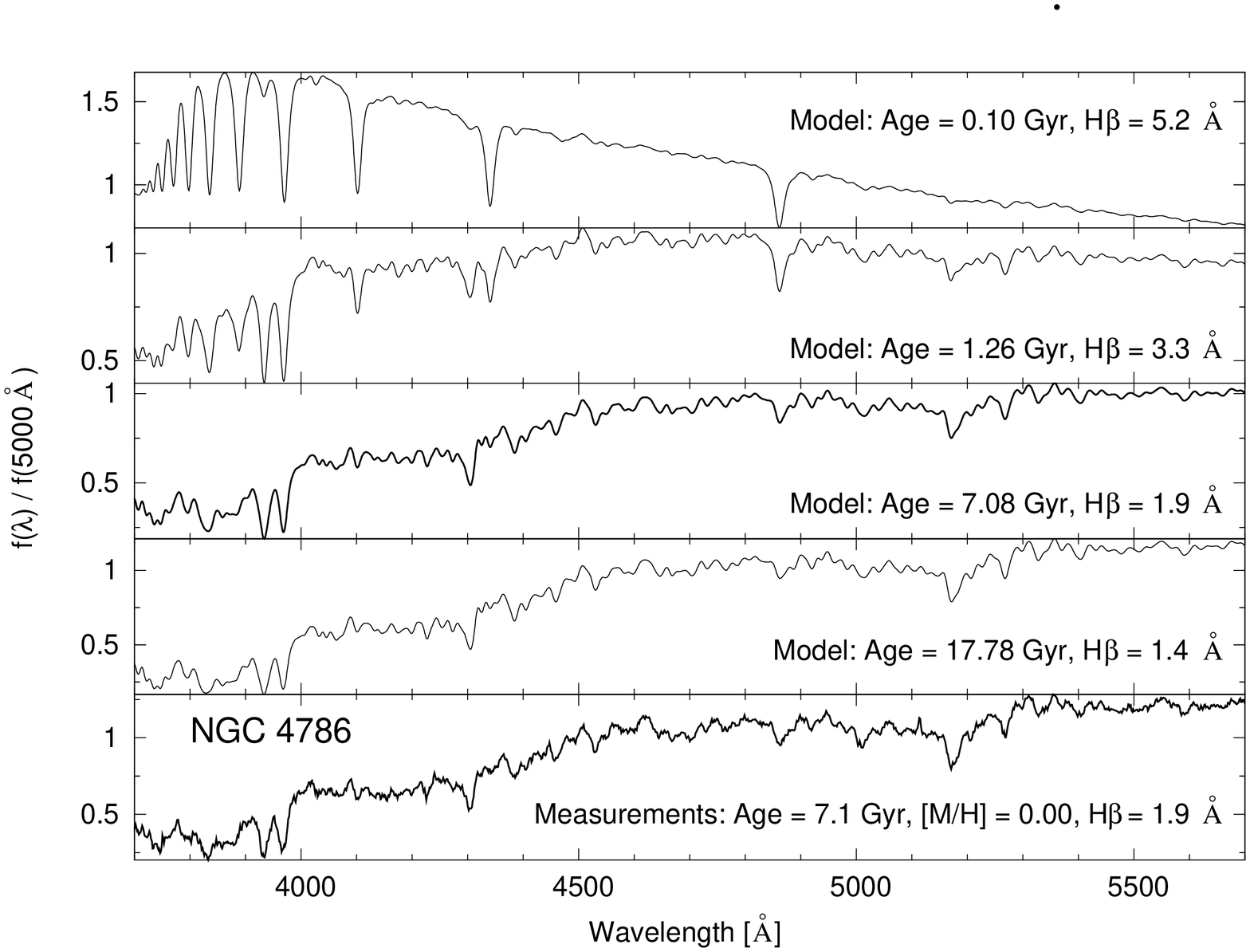}
\caption{A comparison between host-galaxy NGC 4786 with an estimated metallicity around solar and 
several Vazdekis model SEDs of varying ages.  Each model has been broadened to the resolution and 
velocity dispersion of the host spectrum and labeled according to its age and H$\beta$ index measurements.  
The series shows how the models change with age at a fixed metallicity and gives an alternate illustration 
of the technique used to realize our galaxy age/metallicity estimates.  In this case, the age-sensitive index, 
H$\beta$, for NGC 4786 most closely matches that measured for the SED at solar metallicity and a single 
population age of 7.08 Gyr.  \label{example_specs}}
\end{center}
\end{figure}

\begin{figure}
\begin{center}
\includegraphics[angle=270,scale=0.70]{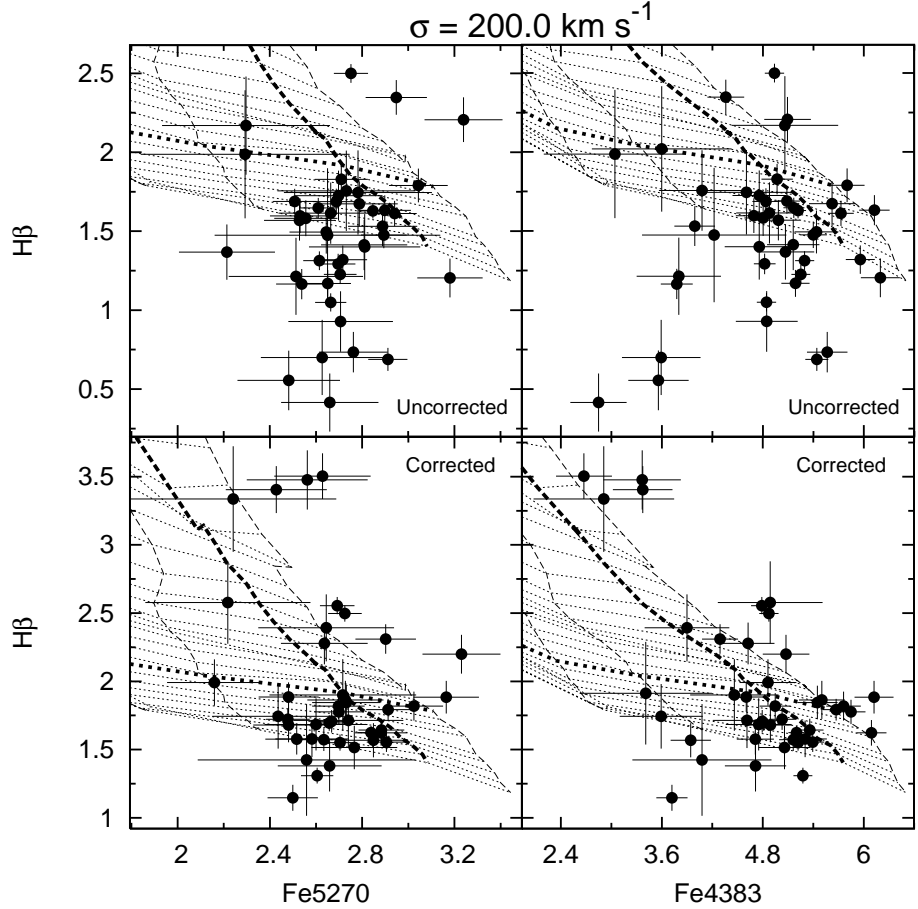}
\caption{Pre/post-emission-corrected diagnostic grids. Age-sensitive absorption-line index equivalent 
width (\AA), H$\beta$, is plotted against the metallicity-sensitive line indices EW, Fe$\lambda$5270 and 
Fe$\lambda$4383, before (top) and after (bottom) emission correction.  The underlying grid reflect the 
Vazdekis model SEDs broadened to a velocity dispersion, $\sigma$ = 200.0 km s$^{-1}$.  Dashed 
lines (near vertical) are models of common metallicity, and dotted lines (near horizontal) are those of 
common age.  The bold dashed line identifies models of solar metallicity, and the bold dotted line marks 
models of age 7.08 Gyr.\label{fig1}}
\end{center}
\end{figure}

A qualitative comparison between several models and a representative host galaxy 
are given in Figure \ref{example_specs}, showing a comparison of the spectrum for host 
galaxy NGC 4786 ($Z \approx Z_{\sun}$) to model SEDs with similar metallicity ($Z \approx Z_{\sun}$) 
but with varying population age.    Quantitative age and metallicity estimates are found through a 
comparison with these model SEDs by way of an interpolation of the index-index diagnostic grids.  
Figure \ref{fig1} shows the Fe$\lambda$5270--H$\beta$ and Fe$\lambda$4383--H$\beta$ grids.  In 
each case the non-emission-corrected host-galaxy indices are plotted in the upper frames and the 
emission-corrected indices are plotted in the lower frames.  Although for each galaxy metallicity and 
age interpolation we broadened the models according to its corresponding velocity dispersion, for the 
sake of presentation we show the whole sample plotted on a model grid broadened to a velocity 
dispersion of $\sigma$ = 200 km s$^{-1}$.    

\clearpage

Computing ages and metallicities for our galaxy samples required the interpolation, and on occasion 
extrapolation, of an irregular grid.  This was accomplished using bivariant polynomial transformations 
of the following form \citep{wol94,card03}:
\begin{eqnarray*}
{{\rm age}}_{i} = \sum_{j}^{N}\sum_{k}^{N-i}{p}_{jk}({indx}_{i})^{j}({indy}_{i})^{k},\\ 
{Z}_{i} = \sum_{j}^{N}\sum_{k}^{N-i}{q}_{jk}({indx}_{i})^{j}({indy}_{i})^{k},
\end{eqnarray*}
where $N$ is the order of the polynomial and \textit{p$_{jk}$} and \textit{q$_{jk}$} are the coefficients 
that are solved for a set of nearest-neighbor points (\textit{indx},\textit{indy}) indexed by \textit{i}.  We 
performed a second-degree approximation, thus requiring us to compute at least 12 coefficients by solving 
two systems of 6 linear equations.  An IDL\footnote{Interactive Data Language: http://www.ittvis.com/idl/idl7.asp .} 
code was written which allowed us to interactively select the nearest 6--12 grid points (the number was 
dependent on the quality of the grid around the host galaxy's grid position) and the coefficients were computed 
by solving the above system of equations using the method of least squares.  

The upper and lower limits for age and metallicity were determined directly from the absorption-line index 
limits outputted by LECTOR.  The extent of the index limits can be seen from the error bars in Figure \ref{fig1}.
The upper and lower limits were treated as unique points for which age and metallicity estimates were made.  In 
the case of large index errors, the grid points chosen by our code were different from those used to determine the 
age/metallicity of the galaxy.  An assumption was made that the upper and lower H$\beta$ index limits 
corresponded to the lower and upper limits, respectively, of our galaxy age estimate. Similarly, the upper 
and lower limit of the Fe$\lambda$5270 index corresponded to the upper and lower limits of the galaxy 
metallicity estimate.  The final uncertainty was recorded as the magnitude of the difference between the age 
and metallicity of the limits and those of the galaxy.  Furthermore, due to the irregular nature of the grid, a 
symmetric error in an index measurement translated into an asymmetric error in age and metallicity.

The results from our analyses are shown in Table \ref{tbl-4}.  Column (1) gives the galaxy name.  Columns 
(2)--(7) give the ages, metallicities, and corresponding errors determined via the Fe$\lambda$5270--H$\beta$ 
diagram, while columns (8)--(13) give the same results ascertained via the Fe$\lambda$4383--H$\beta$ diagram.

\clearpage

\subsection{Field Galaxy Sample}

Aspiring to test the uniqueness of SN~Ia early-type host galaxies relative to random early-type galaxies, we compared the 
global properties of our SN~Ia sample of elliptical host galaxies to those of a general sample of elliptical field 
galaxies from SDSS. The following steps were taken to generate a comparative sample of field galaxies.  The 
SDSS Catalog Archive Server was queried for galaxy spectra within the redshift range 0 $\le z \le 0.15$ that 
had a velocity dispersion measured by SDSS.  The positions of these galaxies were then cross-referenced 
against galaxies in the NASA/IPAC Extragalactic Database (NED), and only those galaxies characterized as 
E/S0 by NED were accepted.  The galaxy spectra of these elliptical field galaxies were then obtained from the 
SDSS Data Archive Server.  The spectra were emission-line corrected using a synthetic Orion spectrum created 
via the IRAF artificial data generation package, ARTDATA.  Furthermore, rather than broadening the models 
to the specific velocity dispersion of each galaxy, we broadened the set of models to velocity dispersions 
in 30 km s$^{-1}$ intervals from 100 km s$^{-1}$ to 340 km s$^{-1}$.  The set of models used in the 
analysis for a given SDSS galaxy spectrum was the set broadened for a velocity dispersion that most closely 
matched that of the galaxy.  Finally, to speed up the calculation, the grid points to perform the diagnostic 
interpolations were automatically chosen for the SDSS sample by an algorithm designed to mimic the 
grid-point selection criteria used for the interactive point selection in the host-galaxy analysis.

\subsection{Data Quality and Consistency Check}

\begin{figure}
\begin{center}
\includegraphics[angle=0,scale=0.50]{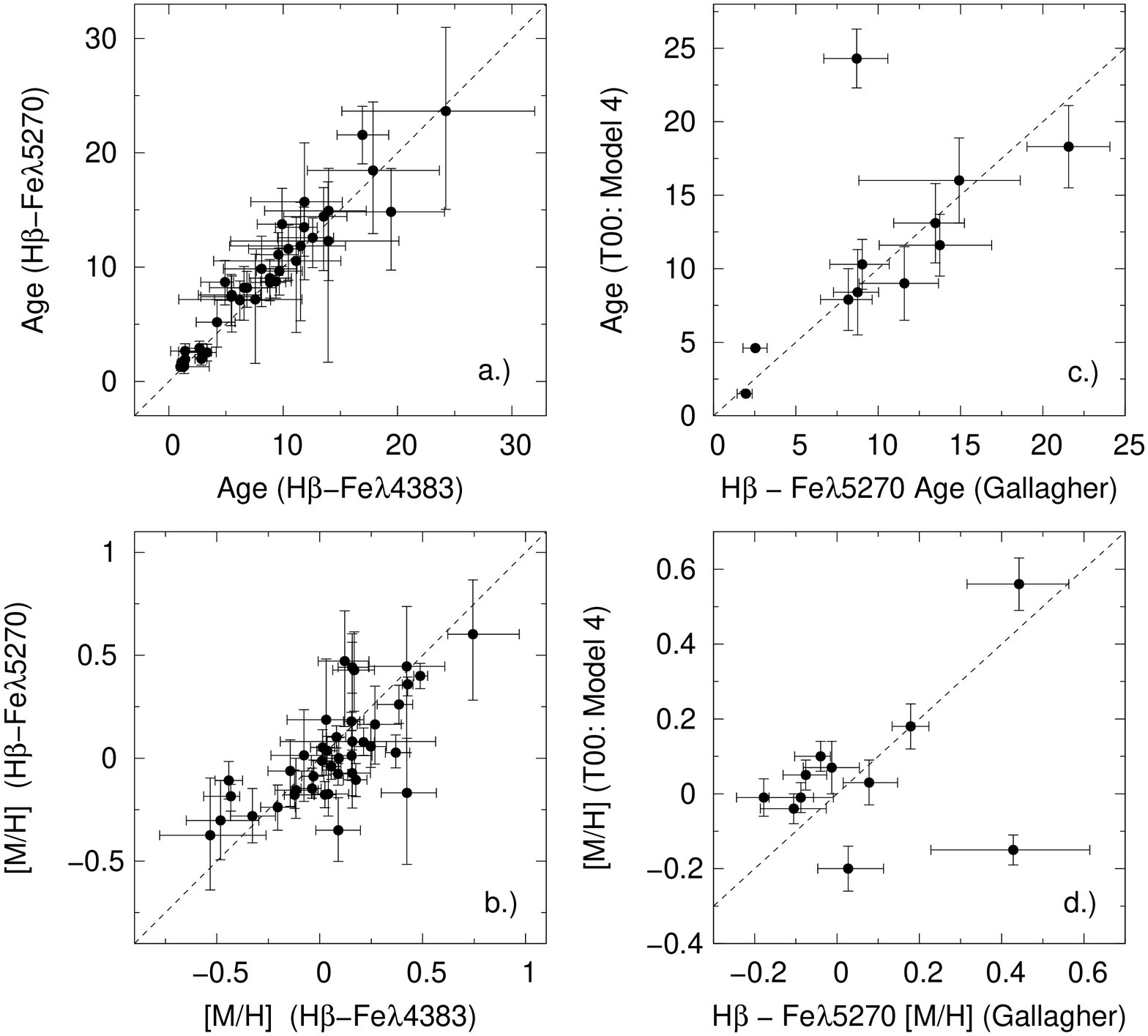}
\caption{Quality and consistency check.  Panels (a) and (b) compare the age and metallicity estimates 
from our two diagnostic grids presented in Figure \ref{fig1}, respectively.  In both cases we find minor 
variations that are nonetheless consistent with unity.  Panels (b) and (d) present a comparison showing 
reasonable agreement between our results and those of \citet{trager00}.\label{fig2}}
\end{center}
\end{figure}

The quality of our data is assessed via three checks, the results of which are shown in Figure \ref{fig2}.
First, we compare the respective age and metallicity estimates obtained from our two index-index diagnostics.  
Figure \ref{fig2}a shows the Fe$\lambda$4383 age estimate plotted against the age estimate of  
Fe$\lambda$5270.  Given that both ages were determined using the same age-sensitive index, namely 
H$\beta$, we would expect these data to be consistent with unity.  As the top-left panel shows, we find 
moderate scatter about unity by virtue of our error estimates.  Nevertheless, the data are consistent with unity, 
with a $\chi$$_{r}^{2}$ of 0.88 for the fit.  Figure \ref{fig2}b shows the metallicity comparison between 
our two index-index diagnostics.  In this case we see more significant dispersion; however, the data are still 
consistent with unity given the calculated uncertainty.  Consequently, as a matter of preference, the galaxy 
ages and metallicities used in the following analysis will be those determined from the H$\beta$--Fe$\lambda$5270 
diagnostic.

Our second quality check is a comparison of our age and metallicity estimates to those of control galaxies 
studied by \citet{trager00} (hereafter T00).  Our data are plotted against the data of T00 in Figure \ref{fig2}c 
and \ref{fig2}d.  T00 made age and metallicity measurements through a central $R_{e}$/2 aperture and we 
are reporting the ``preferred" non-solar abundance ratio (NSAR) model results (model 4).  The age-sensitive 
line index used was H$\beta$, while the two metallicity-sensitive indices were $<$Fe$>$\footnote{$<$Fe$>$ 
$\equiv$ ($\frac{1}{2}$)(Fe$\lambda$5270 + Fe$\lambda$5335).} and Mg~b.  Figure \ref{fig2}c shows the 
age comparison for the two studies with the dotted line shows a model one-to-one relation.  The plot shows 
fairly good agreement between the two studies.  The $\chi$$_{r}^{2}$ of the one-to-one fit is 8.55, though it 
is improved if we throw out NGC 5813 at (8.69, 24.30), reducing $\chi$$_{r}^{2}$ to 2.72.  

The agreement is less encouraging for the metallicity comparison in Figure \ref{fig2}d.  Ignoring the few 
outlier points, the T00 galaxies seem to be systematically more metal rich than those measured in our study.  
This is likely due to the fact that we did not account for NSAR in our analysis.  Studies have shown that the 
abundance ratios in early-type galaxies are often non-solar.  In particular, the Mg/Fe abundance ratio has 
been shown to be larger in more luminous early-type galaxies \citep{o76,p89,w92,v97}.  Given that both 
Mg and Fe lines strongly factored into the metallicity measurements of T00, we should not expect to see a 
straight 1:1 comparative ratio.  Nevertheless, throwing out the points with the two greatest dispersions (NGC 
5813 and NGC 4489), we seem to be measuring the relative metallicities consistently with T00, thereby 
enabling us to find potential trends between age/metallicity and SN~Ia properties. 

We also compare our results to those published in \citet{t05}.  In their study, they seek to set constraints on 
the epochs of early-type galaxy formation through an analysis of 124 early-type galaxies in both high and 
low-density regions.  We have 15 galaxies in common with their study and compare our H$\beta$ index, age, 
and metallicity measurements to those from their analysis.  Similar to the comparison with T00, we see 
general agreement between our respective results.  However, the agreement is far from one to one, with broad 
scatter greater than that seen in our T00 comparison.  Once again, this not entirely unexpected given our lower 
S/N and the fact that we did not accounted for potential $\alpha$ enhancement in our galaxies.  Moreover, they 
were sampling a significantly smaller fraction of the galactic light with an aperture radius of $R_{e}$/10.  Overall, 
our comparisons with both T00 and \citet{t05} show reasonable agreement and confirm the ability of our 
analysis to measure acceptable relative population ages and metallicities for our SN~Ia 
host-galaxy sample. 
     
\begin{figure}[t]
\begin{center}
\includegraphics[angle=0,scale=0.60]{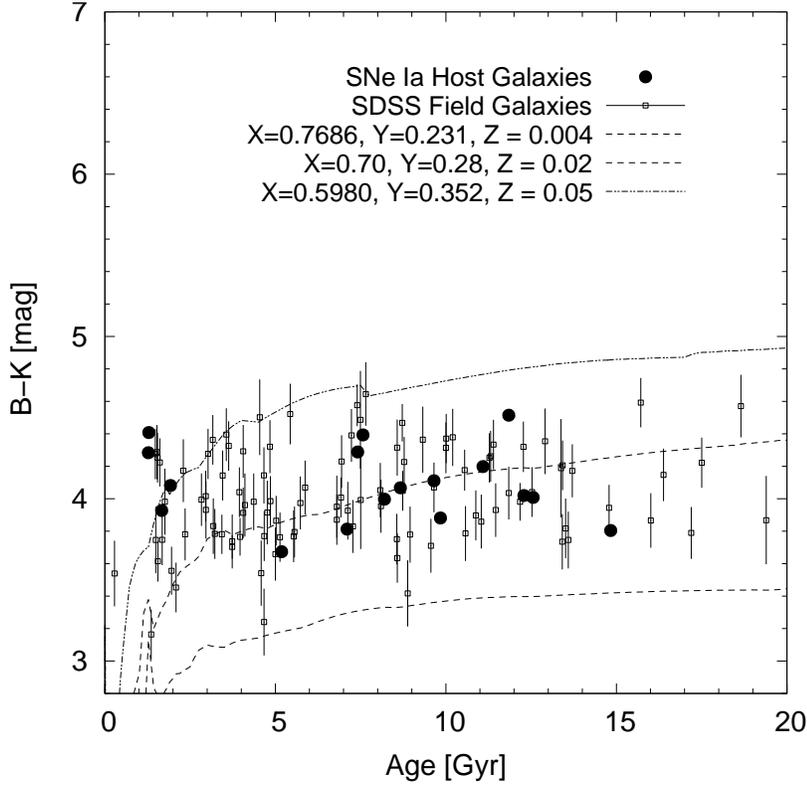}
\caption{The SN Ia host-galaxy $B-K$ color vs. age.  We see that both of the samples fall within the range 
predicted by the evolutionary synthesis models of \citet{bc03}.  The center line is the predicted trend of 
$B-K$ color with age for a population of solar metallicity, the top line is the prediction for a metallicity of 
[M/H] $\approx -0.68$, and the bottom line is the prediction for a metallicity of [M/H] $\approx 0.40$.\label{fig3}}
\end{center}
\end{figure}

Finally, we compare the $B-K$ colors for the host galaxy and SDSS data to the galaxy age; see Figure 
\ref{fig3}.  The expected average $B-K$ color for a sample of elliptical galaxies is 4.0 \citep{j03}, and 
we can see that both our host galaxies and the SDSS galaxies meet this expectation.  The figure also shows 
that both of the samples fall within the range predicted by the evolutionary synthesis models of \citet{bc03}.  
The center line is the predicted trend of $B-K$ color with age for a population of solar metallicity, the top line 
is the prediction for a metallicity of [M/H] $\approx -0.68$, and the bottom line is the prediction for a 
metallicity of [M/H] $\approx 0.40$.

\subsection{Age and Metallicity}

\begin{figure}
\begin{center}
\includegraphics[angle=0,scale=0.65]{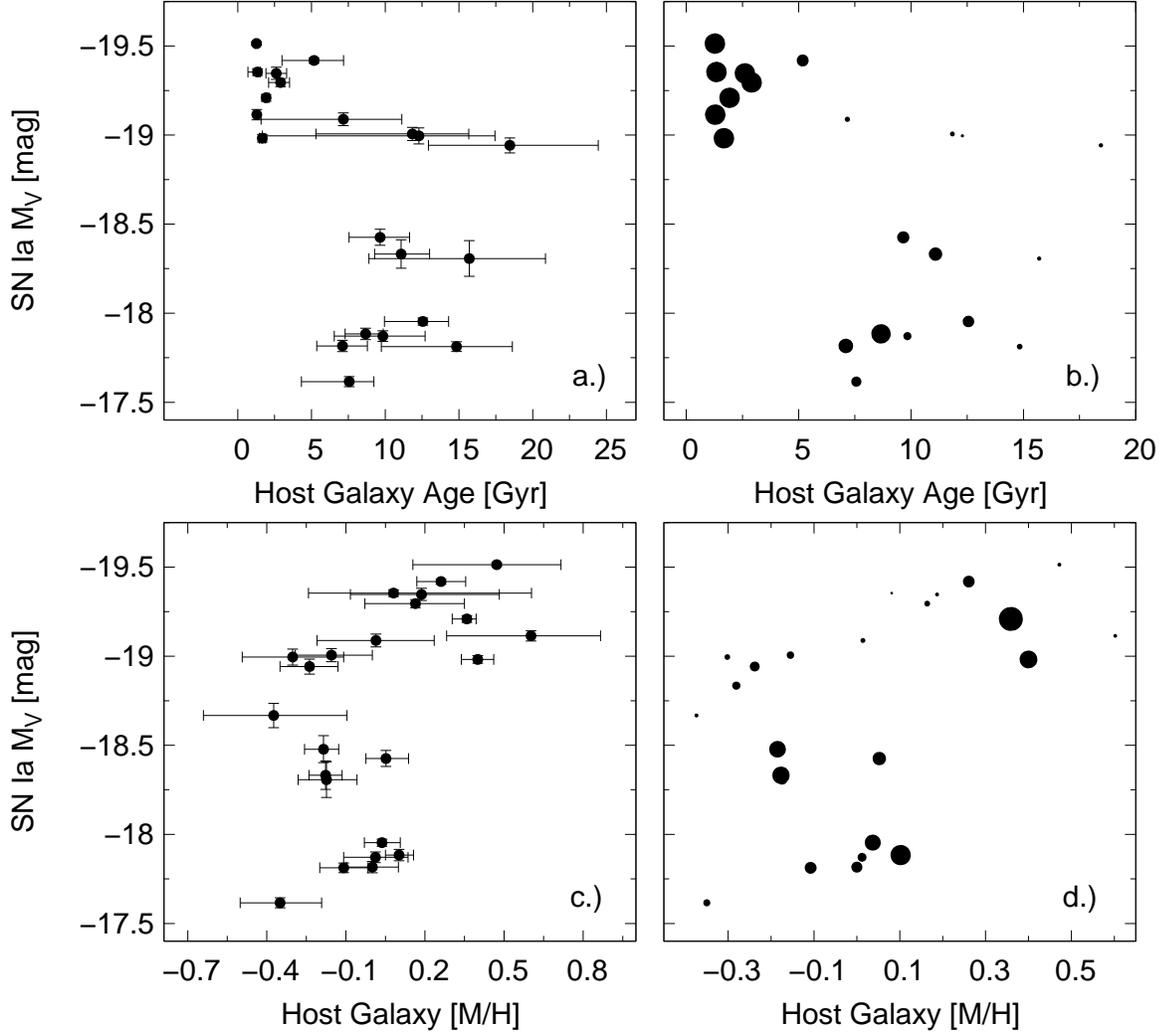}
\caption{The SN Ia peak magnitudes (H$_{0}$ = 72 km$^{-1}$ s$^{-1}$ Mpc$^{-1}$) vs. (a) luminosity-weighted host-galaxy age and (c) metallicity.  Panels (b) and (d) show the same data with the point size reflecting the 
average error along the abscissa.  Point size is inversely proportional to the average error along the abscissa, with the 
exception of our young hosts in (b) where an upper limit is imposed on the point size.  The data in panel (b) 
show a clear trend between SN Ia peak magnitude and host-galaxy age.  The behavior could be evidence of 
either a smooth transition of SN magnitude with age or of two distinct populations of SNe~Ia.  Panel (d) 
shows a less convincing correlation likely arising from the degeneracy between age and metallicity.
\label{fig4}}
\end{center}
\end{figure}

The host-galaxy age and metallicity vs. SN~Ia peak $V$-band magnitude are plotted in Figure \ref{fig4}. 
The peak magnitudes were derived from measurements of the light-curve shape parameter, $\Delta$, fitted using 
MLCS2k2 by the authors of  \citet{jrk07}\footnote{M$_{V}$(peak) = $-19.504 + 0.736\Delta$ + 
0.182$\Delta$$^{2}$ + 5 log (H$_{0}$ / 65 km s$^{-1}$ Mpc$^{-1}$) mag.}.  In Figure \ref{fig4}a, the data show 
that the youngest populations tend to host the brightest SNe~Ia, but with a wide spread.  However, the eye has a 
tendency to gravitate toward large error bars, thereby placing an unwanted amount of attention to those points. 
Consequently, we have replotted the data in Figures \ref{fig4}b and \ref{fig4}d with point sizes that are inversely 
proportional to their uncertainty.  In this way, the data with the smallest uncertainty in age will be represented by 
the largest points. When points with the largest uncertainty are de-emphasized, the trend between age and 
luminosity is very clear, but it is nevertheless difficult to distinguish whether the effect is a smooth transition with 
age or the result of two distinct populations. 

\begin{figure}
\begin{center}
\includegraphics[angle=0,scale=0.6]{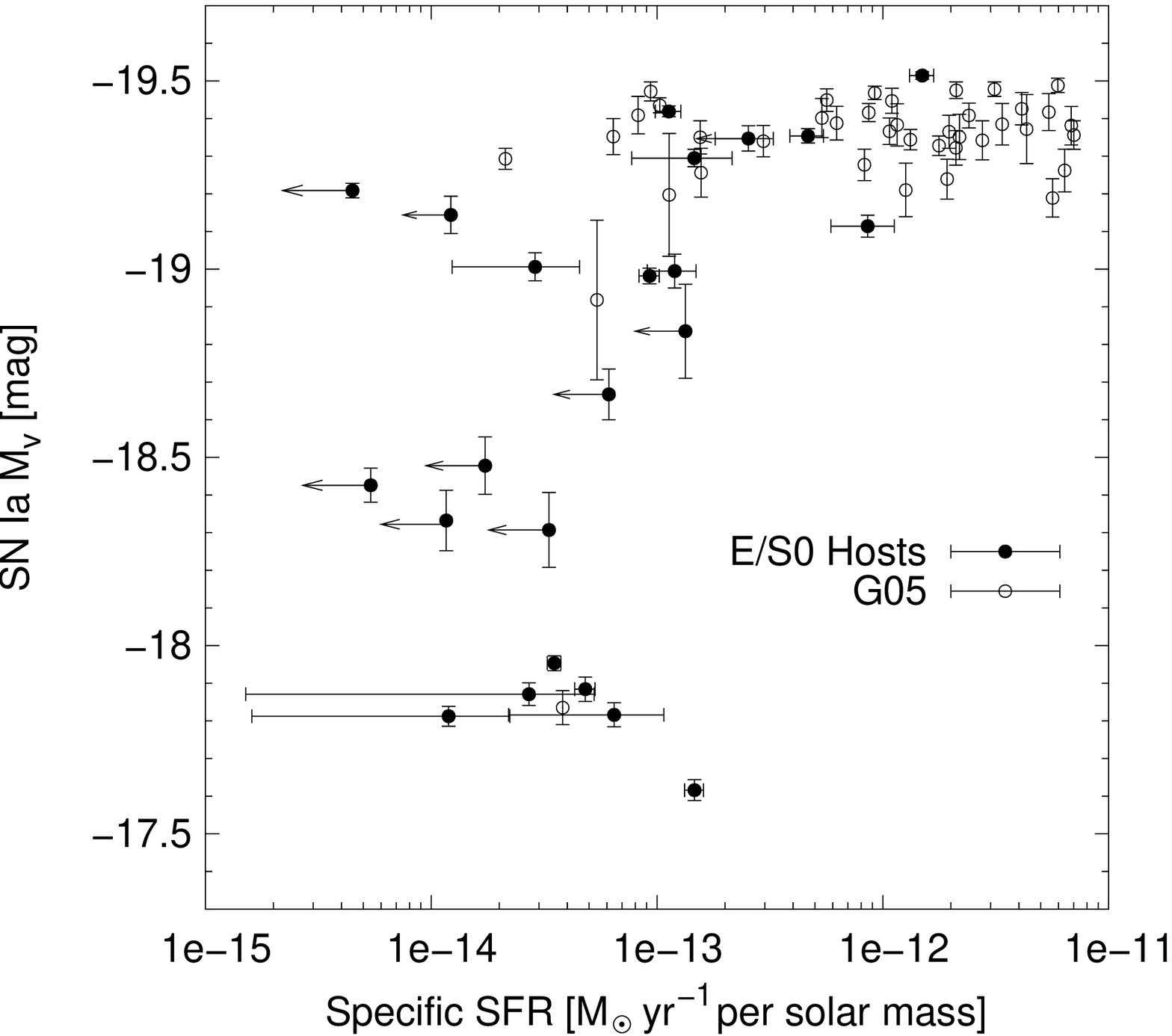}
\caption{The dependence of SN Ia peak magnitude on host-galaxy specific SFR$_{{\rm H}\alpha}$.  Closed 
circles are the current sample of early-type hosts while open circles represent the high-SFR hosts 
of G05.  The points marked by arrows indicate upper limits.  On average, our specific SFR distribution is   
lower than that seen in the sample of high-$z$ ``passive" hosts of  \citet{s06}.  However, aside from the expected 
decrease of SFR over cosmological time, it should be noted that our data only sample half of the galactic 
light given our extraction radius ($R_{e}$), and the SFRs for the ``passive" galaxies of \citet{s06} were 
randomly assigned a rate of $\sim$0.005 M$_{\sun}$ yr$^{-1}$.\label{magsfr}}
\end{center}
\end{figure}

\citet{Gallagher05} found that only luminous SNe~Ia occur in strongly star-forming hosts while E/S0 
galaxies show a wide range of SN~Ia luminosities.  This is confirmed in Figure \ref{magsfr}, where we have
plotted SN~Ia peak magnitude vs. host-galaxy specific star formation, i.e., the 
SFR per stellar mass.  The SFRs were calculated from the H$\alpha$ emission flux using 
the relation of \citet{kennicutt98}, with the distance to each galaxy coming from the SN~Ia luminosities.  
Solid circles represent the host galaxies from the current sample of early-type galaxies, while open circles are from the sample
studied by G05.  We eliminated galaxies from G05 if they were in our current sample since the H$\alpha$ flux 
from G05 was uncorrected for underlying absorption.  The points with arrows represent star formation upper 
limits in which H$\alpha$ emission was buried within the noise.  This trend, combined with Figure \ref{fig4}b, 
suggests that the age of the dominant population in E/S0 galaxies determines the resulting SN~Ia peak 
luminosity.  A correlation between SN~Ia peak luminosity and SFR has also been seen in the high-redshift 
SNLS data \citep{s06}, but their type of search is biased against very low-luminosity events. Our ``young'' 
E/S0 galaxies correspond to their ``passive'' hosts, and the extension to very old populations (with low 
specific SFR) confirms that population age is the major factor that determines the SN~Ia peak luminosity.

\begin{figure}[t]
\begin{center}
\includegraphics[angle=270,scale=0.60]{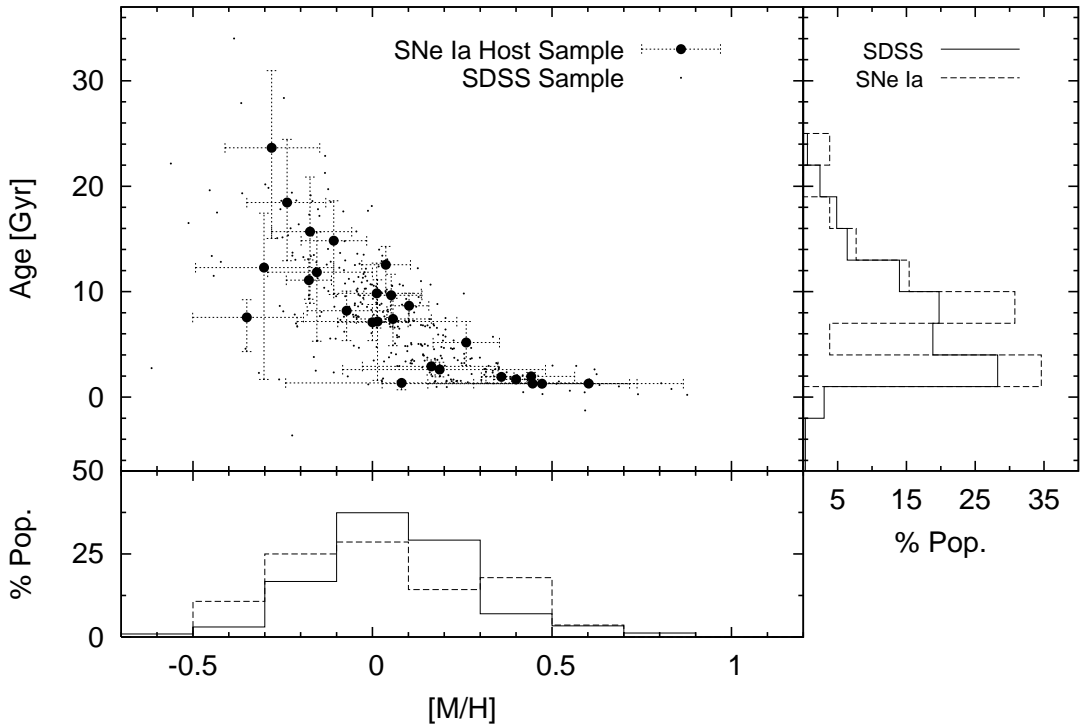}
\caption{An illustration of the age-metallicity degeneracy ``problem" in the study of stellar populations.  The 
plot shows the luminosity weighted age vs. metallicity results for our sample of SN Ia host galaxies and SDSS 
elliptical field galaxies.  The plot reveals the trend stretching from young, metal rich to old, metal poor stellar 
populations.  Histograms are also presented illustrating the age and metallicity distributions for the two 
respective samples.\label{fig5}}
\end{center}
\end{figure}

The plots of host metallicity vs. SN~Ia peak luminosity (Figures \ref{fig4}c and \ref{fig4}d) also show 
a mild correlation. This may be due to the ``age/metallicity degeneracy'' shown in Figure \ref{fig5}. There we 
present measured galaxy age vs. metallicity for both the hosts and SDSS field galaxies, showing that the 
galaxies are not evenly distributed in the diagram but instead concentrated to the lower-left half. This is simply 
due to the evolution of the universe which started metal poor, meaning that there were few galaxies that are both 
old and metal rich and occupy the upper-right of the diagram. Thus, any correlation between age and a supernova 
property will result in some correlation with metallicity as well. Here, the relationship between host age and SN~Ia 
luminosity is so clear that we claim it represents a physical connection. We attribute mild effect in the metallicity 
relation to the lack of galaxies in the lower-right quadrant of Figure \ref{fig4}d, which comes from age correlation 
combined with the fact that there are no old galaxies having high metal abundance in our sample.

\begin{figure}
\begin{center}
\includegraphics[angle=0,scale=0.60]{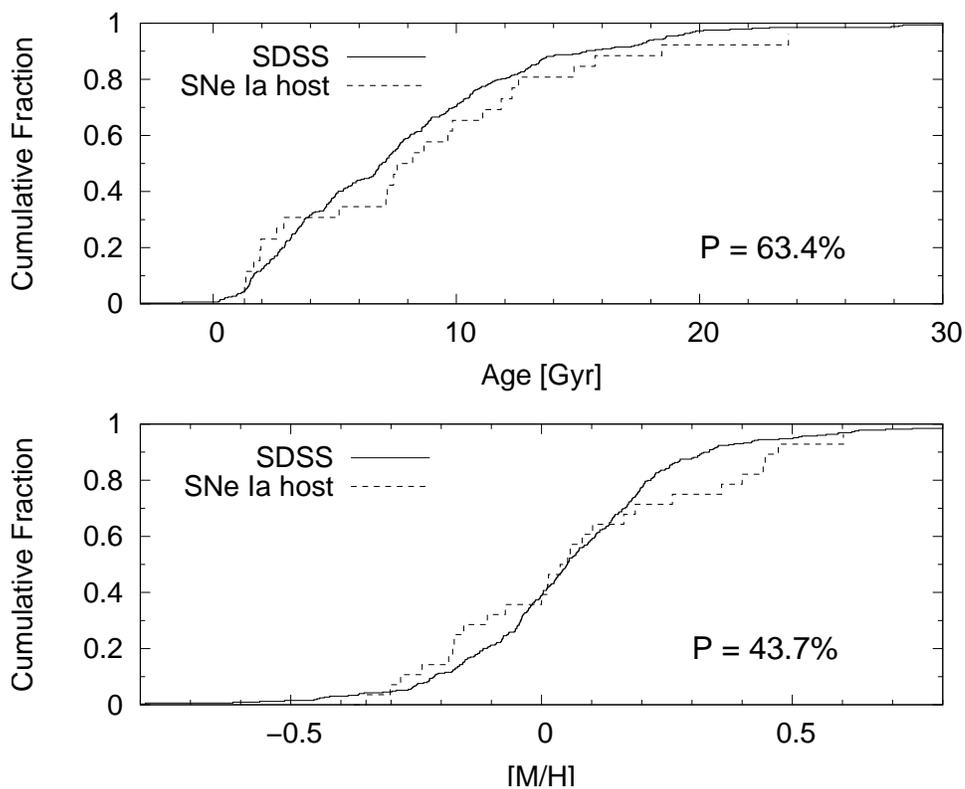}
\caption{Cumulative fraction plots of the SN Ia host galaxy and SDSS galaxy age and metallicity distributions.  
Our sample of SN~Ia host galaxies is clearly similar in both age and metallicity to the SDSS field 
galaxy sample.  We find a 63$\%$ and 44$\%$ probability that the SDSS age and metallicity distributions, 
respectively, represent acceptable parent populations from which the host metallicities were drawn.\label{fig6}}
\end{center}
\end{figure}

The distributions of age and metallicity for the two samples are also shown in Figure \ref{fig5}, with the 
cumulative fraction plots shown in Figure \ref{fig6}.  A Kolmogorov-Smirnov test reveals that there is a 
63\% chance that the host age distribution is drawn from the SDSS field galaxy age distribution, and a 44\% 
chance that the host metallicity distribution is drawn from the field galaxy metallicity distribution.  The results 
suggest that, given the size of the SN Ia sample, the abundance and age of early-type galaxies that 
host SNe~Ia are similar on average to normal early-type field galaxies.  Assuming this, the probability of a 
SN~Ia going off in a given early-type galaxy does not strongly depend on the age or metallicity of the galaxy.

\subsection{Hubble Residuals}

The relation correcting SN~Ia peak luminosity by the light-curve shape reduces the scatter in the 
low-$z$ Hubble diagram to $\sim$ 0.18 mag \citep{jrk07}.  If this residual scatter is purely random, then the 
evolution of age and metallicity in the universe will not induce a bias in cosmological measurements with SNe~Ia.  
However, recent studies such as Timmes, Brown, \& Truran (2003; hereafter TBT03)
and \citet{p06} have predicted that SN calibration methods
should be affected by a cosmic evolution in metallicity.  

In order to investigate this possible effect for SNe~Ia hosted by early-type 
galaxies, we plot our measurements of age and metallicity against post-corrected SN~Ia magnitude residual from the 
best fit to the Hubble diagram.  The Hubble residual (HR) is defined as
\begin{eqnarray*}
{\rm HR} \equiv \mu_{z} - \mu_{SN},
\end{eqnarray*}
where $\mu_{z}$ is the distance modulus determined from the host-galaxy redshift\footnote{Imposed lower 
redshift cutoff at 2500 km s$^{-1}$ to decrease uncertainty due to peculiar velocities.}, and $\mu_{SN}$ is the 
distance modulus determined via the MLCS2k2-corrected SN magnitude.  HR is defined in the conventional 
manner such that overluminous SNe~Ia have a negative Hubble residual\footnote{Note that the opposite 
definition was adopted by \citet{Gallagher05}.}.  The primary component to the uncertainty in HR comes from 
$\Delta{\mu_{SN}}$.  The distance errors estimated from MLCS2k2 were found to be smaller than the scatter 
about the Hubble diagram.  The authors of \citet{jrk07} attributed this to an intrinsic scatter in SN Ia magnitudes
and accordingly imposed a constant additive correction to these distances.  Given that we are probing for 
systematic changes in this intrinsic scatter with age and metallicity, we have not applied this correction to 
$\Delta{\mu_{SN}}$.    

\begin{figure}
\begin{center}
\includegraphics[angle=0,scale=0.55]{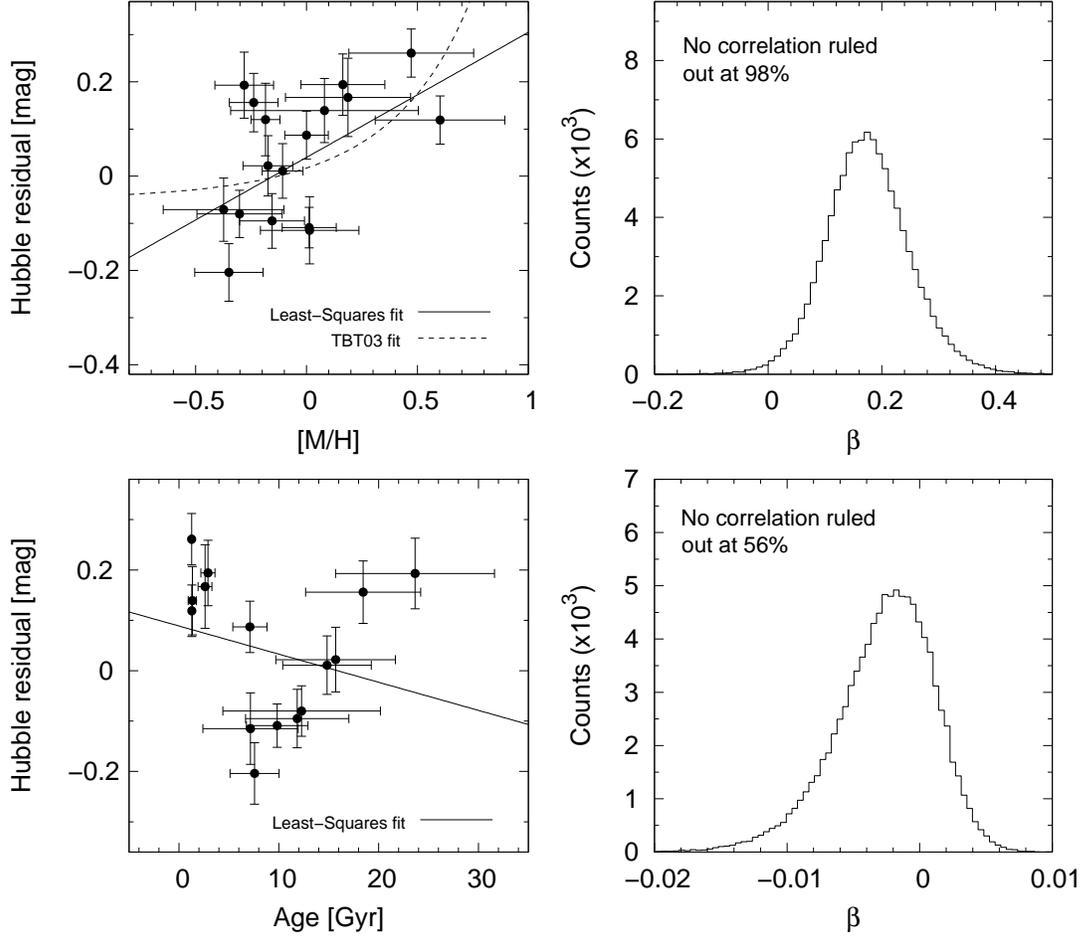}
\caption{SN Ia Hubble residual vs. luminosity-weighted metallicity (top left) and age (bottom left).  
A general trend is found suggesting that more metal rich populations produce intrinsically fainter SNe Ia.  
A least-squares fit finds to the data places a slope for the trend at 0.26.  The dotted line represents the predicted 
trend based on the analytical models of \citet{tbt03}.  The results of our statistical tests are presented in the 
adjacent plots.  We rule out a no-correlation result for the HR vs. metallicity data at the 98\% confidence 
level.  A least-squares fit to the HR vs. age plot yields a nominal slope of $-0.03$ with negligible significance.  
\label{fig7}}
\end{center}
\end{figure}

The top-left panel of Figure \ref{fig7} shows the Hubble residual vs. host-galaxy metallicity along with a 
least-squares fit to the data (solid line). The data and the linear fit show a correlation between host 
metallicity and deviation from the Hubble diagram such that supernovae in metal-rich galaxies are fainter 
than average and metal-poor galaxies host brighter than average events.  The  dashed line is a representation 
of the TBT03 relation (M$_{Ni}$vs$.$ metallicity) showing that, at least qualitatively, the observed trend is 
consistent with that by expected by TBT03.

To map the TBT03 prediction onto the plot, we converted the $^{56}$Ni yields from TBT03 into V-band 
peak magnitude via the models of \citet{hof02} as was done in Gallagher et al. (2005). To avoid an 
over-dependence on models, we also converted Ni masses to peak magnitudes through an empirically 
determined relation using Ni mass estimated by \citet{stritz06} and corresponding V-band luminosities 
given in \citet{jrk07}.  Mapping TBT03 onto Figure \ref{fig7} using this conversion yields a qualitatively
similar trend to both our data and the TBT03-H\"{o}flich model.

%The top-left panel of Figure \ref{fig7} shows the Hubble residual vs. host-galaxy metallicity along with a 
%least-squares fit to the data (solid line).  The dashed line is a representation of the TBT03 relation (M$_{Ni}$ 
%vs$.$ metallicity) showing that at least qualitatively the observed trend is consistent with that by expected by TBT03.  
%To produce this curve we converted the $^{56}$Ni yeilds from TBT03 into SN V-band peak magnitude via the 
%models of \citet{hof02}, thus predicting that metal-rich progentors will produce fainter SN Ia.  One caveat is that 
%inherent in our use of the H\"{o}flich models is a hidden assumption that not only is there a dependence of peak 
%magnitude on M$_{Ni}$ ejected, but that the peak magnitudes within these models are also dependent on the 
%deflagration-to-detonation transition densities.  The potential danger here is that there is no a priori reason for this 
%latter dependence.  However, the impact of this assumption seems to be small given the qualitative purpose of our 
%fitting the TBT03 relation to Figure \ref{fig7}.  Removing our dependence on the H\"{o}flich models we converted 
%our Ni masses to peak magnitudes through an empirically determined relation derived from estimates of Ni mass yields 
%for real SN Ia by \citet{stritz06} and their corresponding V-band luminosities given in \citet{jrk07}.  Fitting TBT03 
%onto Figure \ref{fig7} using this conversion yields a qualitatively similar trend to both the our data and the 
%TBT03-H\"{o}flich model predicting metal-rich progenitors producing fainter SN Ia.    

The top-right panel shows the results of the significance analysis for our linear fit.  For each point 
([M/H]$_{\textit{i}}$,HR$_{\textit{i}}$), the Hubble residual was replaced by a randomly selected HR 
from a Gaussian distribution centered on the actual HR with a standard deviation equal to $\Delta$HR.
The same was done for each metallicity using the corresponding [M/H]$_{\textit{i}}$ and 
$\Delta$[M/H]$_{\textit{i}}$. We performed 100,000 iterations, measured the best-fit slope ($\beta$) to the 
new data each time, and generated the $\beta$ distribution for the test.  The results are seen in the top-right 
panel and show that a no-correlation result ($\beta$ = 0) is ruled out at the 98\% confidence level.  A similar 
correlation between dust/extinction and Hubble residual is not seen for our sample, thereby removing the 
possibility of this correlation being a secondary dust effect.  

Assuming that the least-squares fit to the top-left panel of Figure \ref{fig7} adequately represents the trend of HR with metallicity, 
we would predict a 0.26 mag increase in intrinsic SN~Ia magnitude per unit decline in progenitor (galaxy) 
metallicity.  We have looked at several studies, two theoretical and one observational, that seek to characterize 
cosmic chemical evolution.  The theoretical studies by \citet{co99} and \citet{cen03} predict an approximate
0.05 dex and 0.25 dex drop in metallicity per unit redshift, respectively.  The observational study 
conducted by \citet{kk07} find a preliminary result of 0.15 dex/$z$.  Taking the average of these three studies 
(0.15 dex/$z$) for the true decline in metallicity with redshift would translate into a 0.039 mag decrease in 
intrinsic SN Ia magnitude per unit redshift.  Assuming an ideal SN Ia data set out to $z < 1.8$,  
a $\chi$$^{2}$ minimization of three parameters ($\Omega$$_{M}$, $w$, and H$_{0}$\footnote{Flat geometry 
assumed.}) applied to a set of model SNe~Ia subject to this 4\% systematic error would induce an approximate 
9\% systematic error on the measurement of the equation-of-state parameter $w = P/(\rho c^2)$.

\citet{jrk07} showed that the scatter of SN magnitudes about the Hubble diagram is lower in E/S0 galaxies 
($\sim$ 0.13 mag) than for sample as a whole ($\sim$ 0.18 mag).  Our sample of E/S0 hosts show a scatter 
of approximately 0.14 mag.  By applying a correction to SN magnitudes consistent with our least-squares
fit, we reduce the scatter about the Hubble diagram to approximately 0.11 mag, illustrating the potential for 
such a correction to significantly improve SN~Ia distance measurements.  
 
The bottom-left panel of Figure \ref{fig7} shows the age of the parent galaxy vs. the Hubble residual of the SN Ia.  The plot
does not reveal a significant trend of HR with host-galaxy age.  Indeed, the probability of a zero-correlation 
result is only ruled out at the 56\% confidence level, suggesting that SN Ia magnitudes corrected for light-curve 
shape are likewise being corrected for any age bias.  It should be noted that two galaxies (CGCG 016-058 and 
MCG+07-41-001) were removed from the age plot because their positions on their respective diagnostic grids 
required a difficult extrapolation that rendered their age estimates extremely uncertain.  

\subsection{SN Ia Rate vs. Population Age}

Since the late 1970s, observations have shown that SNe Ia are more prevalent in star-forming late-type galaxies 
than in early-type galaxies \citep{ot79}.  This fact has been confirmed again and again with new studies showing 
that the SN~Ia rate per unit mass is significantly higher in blue/late-type galaxies than in red/early-type galaxies 
\citep{van90,dvl94,m05,Leaman08}.  The current explanation for these observations is that there are ``prompt" 
and ``delayed" (``tardy'') SN~Ia explosions. The prompt component is dependent on the rate of recent star formation, and 
the delayed component is dependent on the total number of low-mass stars.  The combination of these two 
components is believed to form the overall observed SN Ia rate \citep{sb05,m05,s06,Leaman08}.  

\begin{figure}[t]
\begin{center}
\includegraphics[angle=0,scale=0.75]{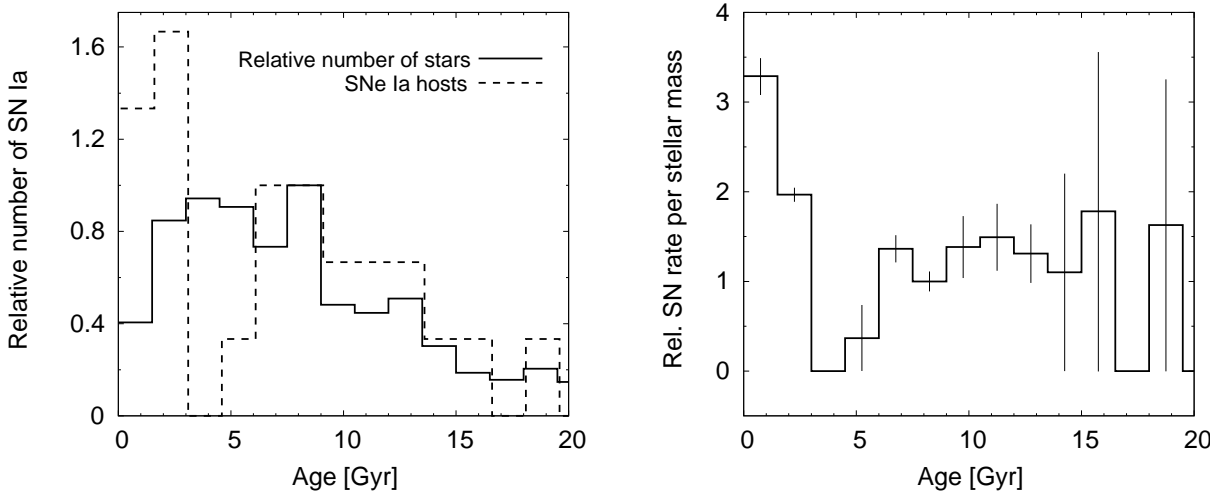}
\caption{The relative SN Ia rate per stellar mass as a function of age in early-type galaxies.  Stellar masses 
were approximated from each galaxy's $K$-band luminosity.  The left panel shows numbers of host galaxies, 
or SNe, per age bin along with the expected number of stars per age bin derived using the SDSS elliptical 
field galaxies.  The right panel gives the expected relative SN Ia rate per age bin, showing a consistency with 
the ``dual-component" model.
\label{fig8}}
\end{center}
\end{figure}

\begin{figure}[t]
\begin{center}
\includegraphics[angle=270,scale=0.60]{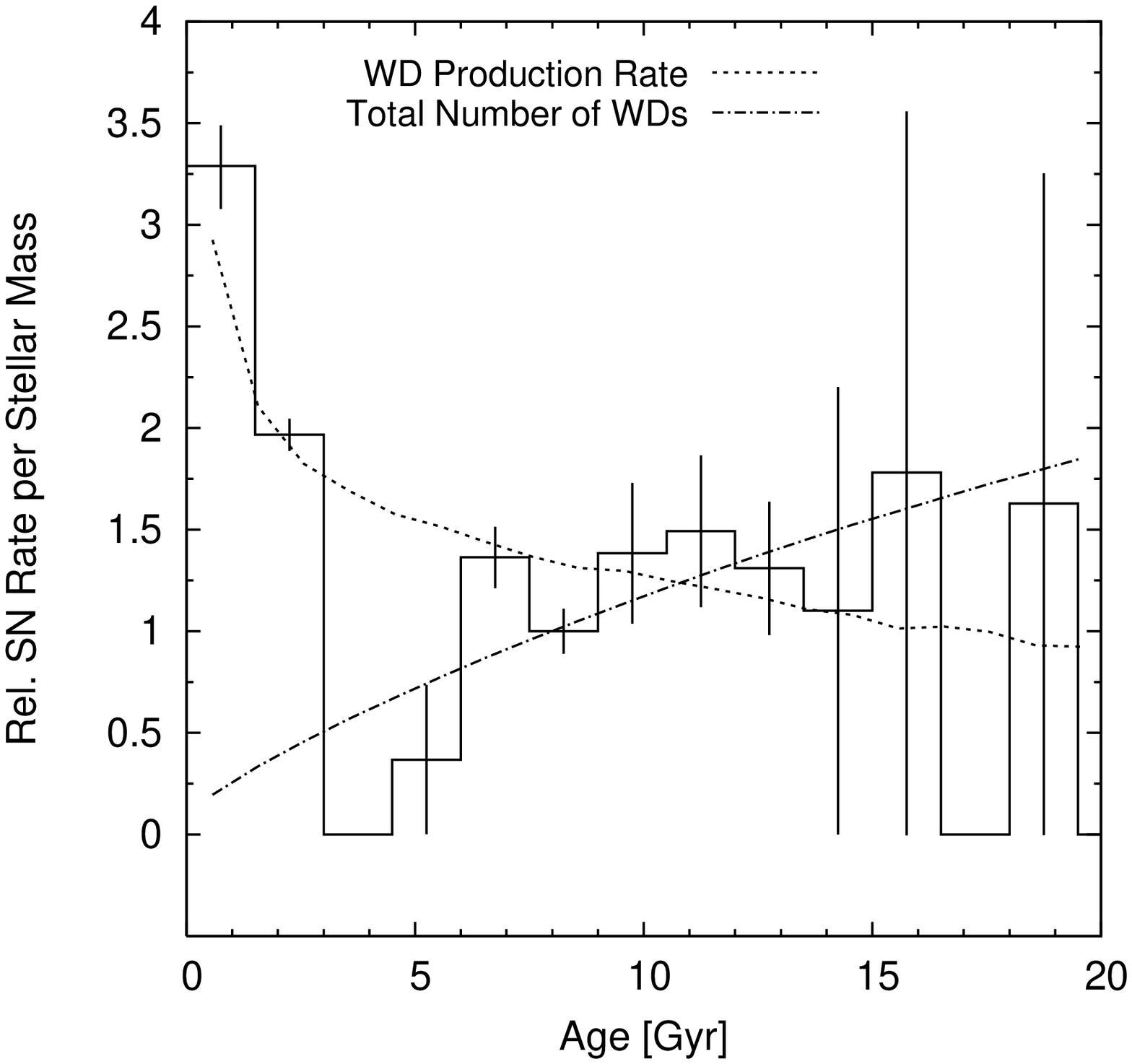}
\caption{The relative SN Ia rate per stellar mass is replotted, showing the bimodal nature of the distribution.  
The dotted line represents the rate of white dwarf production while the dash-dot line is the accumulated number 
of white dwarfs as a function of time since the burst.\label{fig9}}
\end{center}
\end{figure}

\citet{s06} investigated the parameters shaping the overall SN~Ia rate by observing the relative presence 
of SNe~Ia in high-$z$ host galaxies of differing mass and SFR.  Our direct age measurement means 
that we can do a similar analysis at low redshifts by estimating the relative SN~Ia rate as a function of host-galaxy 
age.  Figures \ref{fig8} and \ref{fig9} summarize the details of our analysis on the relative SN~Ia rate in 
elliptical host galaxies of various ages.  

The panel on the left of Figure \ref{fig8} contains two unique histograms.  The dotted-line histogram represents the 
number of host galaxies, or equivalently the number of SNe~Ia, per age bin (identical to the distribution in 
Figure \ref{fig5} with decreased bin size).  The solid-line histogram is the expected number of stars per age bin.  
Each of the distributions has been normalized such that the bin encompassing 8 Gyr contains exactly one element.  
The number of stars is derived from the $K$-band SDSS galaxy distribution, the $K$ band being a relatively accurate 
tracer of stellar mass.  $K$-band apparent magnitudes for the SDSS sample were compiled from NED and the 
distances to the galaxies in the sample were calculated from the SDSS redshift measurements, where care was taken 
to exclude those galaxies with $v_{r} < 2500$ km s$^{-1}$.  The $K$-band absolute magnitudes, $M_{g}$, were 
derived from the distances and apparent magnitudes.  Finally, the number of stars per galaxy was approximated by 
the expression
\begin{equation}
N_{stars} = 100^{[(M_{K,\sun} - M_{g}) / 5.0]}.
\end{equation}   
The variable $M_{K,\sun}$ is the corresponding absolute magnitude of the Sun in the $K$ 
band\footnote{$M_{K,\sun}$ = 3.28.}.  The number of stars per age bin was then calculated by summing 
the stars in each galaxy in each age bin.

The relative SN~Ia rate per unit stellar mass is defined as the ratio of the number of SNe per age bin to the 
expected number of stars per age bin.  The panel on the right of Figure \ref{fig8} shows the expected SN~Ia rate 
per unit stellar mass as a function of age. The plot predicts a rate that is high for young galaxies, falls 
for intermediate-age galaxies, and increases and moderates for older populations.  This result is consistent 
with the ``dual-component model,'' given that the SN rate at low age could arise from a short delay time 
population of SNe~Ia predicted to occur in galaxies with high SFR, and the rise in SN rate at late age could likely  
result from the predicted rise in SN rate with stellar mass \citep{s06}.  However, we take the less general 
route and suggest the possibility that the rate at low population age is proportional to the rate of WD 
production following an initial burst of star formation.  This function can be seen in Figure \ref{fig9}.  Similarly, 
the late-time SN rate can be traced by the cumulative number of WDs present in the galaxy (Figure \ref{fig9}).  
This value is determined for an age, $t = t_{0}$, by integrating the WD production rate from time $t = 0$ to $t = 
t_{0}$.  These components of the SN~Ia rate are related to those of \citet{s06}, and yet they specifically target the 
population of stars believed to be the progenitors of SN~Ia explosions.  

An important caveat to this analysis is that 
we are only sampling the dominant stellar population within a given host galaxy.  It has been shown that a significant 
portion (15--30\%) of elliptical and lenticular (S0) galaxies show evidence of recent star formation \citep{y05,k07}, which 
has the potential to compromise our conclusion that SNe~Ia exploding in old stellar populations possess longer delay 
times.  

\section{Summary and Conclusions}

We have analyzed a sample of early-type galaxies that have hosted SNe~Ia.  
Comparing the data with stellar population synthesis models, we have measured the global 
age and metallicity of the SN~Ia elliptical host galaxies and a general sample of elliptical 
galaxies from the SDSS.  Our results indicate that there is likely a significant correlation 
between age or metallicity and SN~Ia absolute magnitude.  Given the mounting evidence 
of a SN Ia rate dependence on specific SFR, we find it most likely that the SN~Ia peak 
magnitudes are correlated with age, and the observed trend with metallicity is merely an 
artifact brought about by the evolutionary entanglement of age and metallicity.

We find the global distributions of age and metallicity to be similar to those of our SDSS elliptical galaxy 
sample, suggesting that the presence, or absence, of a SN~Ia in an elliptical galaxy is not dependent on 
either age or metallicity.  Moreover, we detect a trend between early-type host-galaxy metallicity and the 
residuals from the Hubble diagram at the 98\% confidence level.  This trend is consistent with the predictions 
of TBT03 and with a trend observed for late-type galaxies by \citet{Gallagher05}; it suggests that metal-rich 
galaxies produce underluminous SNe~Ia, even after correcting for their light-curve shapes.  Furthermore, we 
conclude that the failure to apply a metallicity correction to SN~Ia magnitudes could potentially introduce a 
9\% error into current and future measurements of $w$. 

Finally, we find the predicted SN~Ia rate as a function of age for our sample of early-type galaxies, and 
determine it to be moderately consistent with the ``dual-component model'' of \citet{s06}.  Also, we 
describe our preferred model governing the SN~Ia rate in which the ``prompt" component is proportional 
to the WD production rate and the ``delayed" component scales with the cumulative WD population.  

\acknowledgments

Partial funding for this work came through NASA LTSA grant NAG5-9364.
Supernova research at Harvard University is supported by NSF Grant AST06-06772.
A.V.F.'s supernova group at the University of California, Berkeley is supported by
NSF grant AST--0607485, as well as by the TABASGO
Foundation.  KAIT was made possible by generous donations from Sun
Microsystems, Inc., the Hewlett-Packard Company, AutoScope Corporation, Lick
Observatory, the NSF, the University of California, and
the Sylvia and Jim Katzman Foundation.  We thank Sumner Starrfield and the Arizona 
State University for granting us access to their computer network that proved 
invaluable for the completion of this work.

\begin{deluxetable}{lcccccc}
\tabletypesize{\small}
\tablecolumns{7}
\tablewidth{0pc}
\tablecaption{Absorption-Line Galaxy Sample\label{tbl-1}}
\tablehead{
\colhead{Galaxy} & \colhead{SN} & \colhead{$M_{V}$}& \colhead{Morphology} & 
\colhead{PA $($$^{\circ}$$)$} & \colhead{Ap. Width ($\arcsec$)} &
\colhead{L.C. ref.\tablenotemark{\dagger}} \cr 
\colhead{(1)}&\colhead{(2)}&\colhead{(3)}&\colhead{(4)}&\colhead{(5)}&
\colhead{(6)} &\colhead{(7)}
}
\startdata
\cutinhead{SN Ia host-galaxy Sample}
NGC 4374     &       1991bg  & -17.88 &      E1      &       90      &       9.63    &       1		\\
NGC 4526     &       1994D   & -18.98 &      S0      &       110     &       10.78   &       1		\\
NGC 4493     &       1994M   & -19.09 &      E       &       40      &       4.03    &       1 		\\
CGCG 016-058 &      1994T   & -18.67 &      Sa      &       -30     &       3.01       &      1       	\\
NGC 2962     &       1995D   & -19.42 &      S0+     &       0       &       6.90    &       1		\\
NGC 5061     &       1996X   & -19.21 &      Sa      &       90      &       4.17    &       1		\\
NGC 5005     &       1996ai  & -19.58 &     Sbc     &       -30     &       12.66   &       1		\\
NGC 5308     &       1996bk  & -18.33 &      S0-     &       60      &       10.78   &       1		\\
NGC 5490     &       1997cn  & -17.87  &       E     &       0       &       5.17    &       1		\\
NGC 5440     &       1998D   & \nodata &     Sa      &       40      &       6.19    &       \nodata 	\\
UGC 11149    &       1998dx  & -19.14 &     Sab     &       -40     &       6.75    &       1		\\
NGC 7131     &       1998co   & \nodata &    S0      &       -35     &       4.45    &       1		\\
NGC 6411     &       1999da   & -17.81 &     E       &       70      &       8.19    &       1		\\
NGC 2841     &       1999by  & -17.95 &      Sab     &       50      &       5.88    &       1 		\\ 
NGC 6038     &       1999cc  & -19.00 &      Sc      &       105     &       8.19    &       1		\\
NGC 2986     &       1999gh  & -18.43 &      E2;HII  &       10      &       6.90    &       1		\\
UGC 11198    &       2000dm  & -18.94 &      Sab     &       45      &       7.06    &       2		\\
CGCG 189-024 &       2002G   &  -19.347 &    E       &       10      &       4.16    &       2		\\
UGC 04322    &       2002he  & -19.01 &      E       &       40      &       5.32    &       2		\\
CGCG 141-044 &      2001bf  & -19.514 &     \nodata &       100     &       5.04       &      2       	\\
Anon	     &       2002aw & -19.35 &       Sb (f)  &       80      &       13.94     &     2       	\\
NGC 4786     &       2002cf  & -17.82 &      E+pec   &       -20     &       6.47    &       2		\\
NGC 6702     &       2002cs  & \nodata &     E       &       55      &       6.31    &       \nodata 	\\
MCG+07-41-001& 	2002do  & -18.48 &       E1      &       0       &       20.85   &       2		\\
NGC 7761     &       2002ef  & -19.30 &      S0      &       90      &       4.61    &       2		\\
NGC 6986     &       2002el  & -19.11 &      SB0-    &       5       &       5.89    &       2		\\
MCG-01-25-009   &       2003D   & -18.31 &      E1      &       10      &       5.18    &       2		\\
IC 0522 	&       2003Y   & \nodata &     S0      &       -20     &       7.34     &       \nodata 	\\
UGC 03787       &       2003ch  & \nodata &     E-S0    &       0       &       4.47    &       \nodata 	\\
NGC 4059        &       2005bl  & -17.62 &     E       &       40      &       4.61    &       3	\\
\cutinhead{\citet{trager00} Comparison Sample}
NGC 3608        &       \nodata  & \nodata &    E2      &       80      &       6.19    &       \nodata 	\\
NGC 4472        &       \nodata  & \nodata &    E2/S0   &       -20     &       12.80   &       \nodata 	\\
NGC 4478        &       \nodata  & \nodata &    E2      &       40      &       6.17    &       \nodata 	\\
NGC 4489        &       \nodata  & \nodata &    E       &       110     &       4.61    &       \nodata 	\\
NGC 4552        &       \nodata  & \nodata &    E;HII   &       -30     &       6.31    &       \nodata 	\\
NGC 4649        &       \nodata  & \nodata &    E2      &       90      &       13.09   &       \nodata 	\\
NGC 5638        &       \nodata  & \nodata &    E1      &       -20     &       8.04    &       \nodata 	\\
NGC 5813        &       \nodata  & \nodata &    E1-2    &       -20     &       6.17    &       \nodata 	\\
NGC 5846        &       \nodata  & \nodata &    E0-1    &       50      &       11.35   &       \nodata 	\\
NGC 6127        &       \nodata  &\nodata &     E       &       90      &       5.32    &       \nodata 	\\
NGC 6703        &       \nodata  & \nodata &    SA0-    &       90      &       5.17    &       \nodata 	\\
\enddata
\tablenotetext{\dagger}{(1) \citet{jrk07}; (2) \citet{gan08,fili01,fili05a}; (3) \citet{tau08,g05a}.}
\end{deluxetable}
\clearpage

\begin{deluxetable}{lccc}
\tabletypesize{\footnotesize}
\tablecaption{Cross-Correlation Results\label{tbl-2}}
\tablewidth{0pt}
\tablehead{\colhead{Galaxy} & \colhead{$v_{r}$ (km s$^{-1}$)} & 
\colhead{$\delta v_{r}$ (km s$^{-1}$)} & \colhead{$\sigma$ (km s$^{-1}$)} \cr 
\colhead{(1)}&\colhead{(2)}&\colhead{(3)}&\colhead{(4)}
}
\startdata
Anon	        & 7932.43	& 19.66 & 136.91 \\
CGCG 016-058	& 10411.31	& 17.79 & 147.99 \\
CGCG 141-044	& 4665.13	& 23.01 & 109.73 \\
CGCG 189-024	& 10113.14	& 21.01 & 188.70 \\
IC 0522		& 5092.26	& 19.05 & 170.76 \\
MCG+07-41-001	& 4584.86	& 68.68 & 344.72 \\
MCG-01-25-009	& 6623.73	& 21.58 & 251.97 \\
NGC 2841	& 633.49	& 25.43 & 244.72 \\
NGC 2962	& 1967.99	& 19.01 & 186.61 \\
NGC 2986	& 2310.81	& 33.56 & 283.67 \\
NGC 3608	& 1237.80	& 21.24 & 209.54 \\
NGC 4059	& 7195.99	& 33.21 & 210.88 \\
NGC 4374	& 1020.48	& 30.57 & 316.56 \\
NGC 4472	& 957.20	& 30.95 & 309.86 \\
NGC 4478	& 1352.18	& 17.67 & 183.09 \\
NGC 4489	& 948.10	& 11.75 & 106.89 \\
NGC 4493	& 6957.26	& 20.98 & 214.39 \\
NGC 4526	& 604.65	& 24.90 & 235.75 \\
NGC 4552	& 316.64	& 27.18 & 282.82 \\
NGC 4649	& 1087.03	& 37.32 & 344.45 \\
NGC 4786	& 4611.97	& 33.49 & 288.86 \\
NGC 5005	& 932.64	& 23.29 & 207.73 \\
NGC 5061	& 2059.50	& 22.15 & 218.96 \\
NGC 5308	& 2014.72	& 22.08 & 241.36 \\
NGC 5440	& 3697.03	& 27.35 & 238.53 \\
NGC 5490	& 4965.75	& 37.73 & 338.21 \\
NGC 5638	& 1640.05	& 20.15 & 172.96 \\
NGC 5813	& 1948.27	& 28.57 & 248.58 \\
NGC 5846	& 1692.00	& 27.71 & 238.42 \\
NGC 6038	& 9399.72	& 32.83 & 183.91 \\
NGC 6127	& 4715.52	& 29.99 & 264.50 \\
NGC 6411	& 3726.29	& 23.52 & 186.15 \\
NGC 6702	& 4721.40	& 32.71 & 209.44 \\
NGC 6703	& 2384.69	& 20.62 & 199.69 \\
NGC 6986	& 8534.95	& 25.52 & 267.16 \\
NGC 7131	& 5409.40	& 36.77 & 185.25 \\
NGC 7761	& 7192.42	& 19.90 & 211.39 \\
UGC 11149	& 16137.64	& 92.35 & 146.10 \\
UGC 11198	& 4518.02	& 19.91 & 145.28 \\
UGC 03787	& 8612.81	& 57.54 & 140.66 \\
UGC 04322	& 7376.56	& 23.98 & 248.08 \\
\enddata
\end{deluxetable}
\clearpage

\begin{deluxetable}{lccc}
\tabletypesize{\small}
\rotate
\tablecaption{Lick/IDS Indices for the Study\label{tbl-3}}
\tablewidth{0pt}
\tablehead{
\colhead{} & 	\colhead{Index Bandpass} & \colhead{Pseudocontinua}  & \colhead{} \cr
\colhead{Name} & \colhead{(\AA)} & \colhead{(\AA)}  & \colhead{Species Measured}
}
\startdata
H$\beta$ & 4847.875--4876.625 & 4827.875--4847.875, 4876.625--4891.625 & H, (Mg), (Cr), C \\
Fe5270 & 5245.650--5285.650 & 5233.150--5248.150, 5285.650--5318.150 & Fe, C, (Mg), Ca \\
Fe4383 & 4369.125--4420.375  & 4359.125--4370.375, 4442.875--4455.375 & Fe, C, (Si) \\
\enddata
\end{deluxetable}
\clearpage

\begin{deluxetable}{lcccccccccccc}
\tabletypesize{\scriptsize}
\rotate
\tablecaption{Ages and Metallicities\label{tbl-4}}
\tablewidth{0pt}
\tablehead{
\colhead{Galaxy} & \colhead{Fe$\lambda$5270} & \colhead{}  & \colhead{} & \colhead{}  & \colhead{} & \colhead{} &
\colhead{Fe$\lambda$4383} & \colhead{}  & \colhead{} & \colhead{}& \colhead{}& \colhead{} \cr  
\colhead{} & \colhead{[M/H]} & \colhead{+$\delta$} & \colhead{$-\delta$} & \colhead{Age (Gyr)} & \colhead{+$\delta$}  & \colhead{$-\delta$} &
\colhead{[M/H]} & \colhead{+$\delta$} & \colhead{$-\delta$} & \colhead{Age (Gyr)} & \colhead{+$\delta$}  & \colhead{$-\delta$}  \cr  
\colhead{(1)}&\colhead{(2)}&\colhead{(3)}&\colhead{(4)}&\colhead{(5)}&\colhead{(6)}&\colhead{(7)}&
\colhead{(8)}&\colhead{(9)}&\colhead{(10)}&\colhead{(11)}&\colhead{(12)}&\colhead{(13)}
}
\startdata
Anon	         & 0.081	  & 0.523	  & 0.323	  & 1.338	  & 0.194	  & 0.644	  & 0.158	  & 0.405	  & 0.350	  & 1.338	  & 0.150	 & 0.644	\\
CGCG 016-058	 & -0.374	  & 0.278	  & 0.266	  & 23.454	  & 13.895	  & 17.898	  & -0.532	  & 0.271	  & 0.246	  & 26.083	  & 15.953	 & 22.769	\\
CGCG 141-044	 & 0.472	  & 0.244	  & 0.318	  & 1.273	  & 0.101	  & 0.101	  & 0.121	  & 0.118	  & 0.129	  & 1.256	  & 0.078	 & 0.061	\\
CGCG 189-024	 & 0.187	  & 0.295	  & 0.270	  & 2.611	  & 0.700	  & 0.694	  & 0.031	  & 0.183	  & 0.190	  & 2.891	  & 0.829	 & 2.027	\\
IC 0522	          & -0.072	  & 0.112	  & 0.110	  & 8.200	  & 1.858	  & 2.849	  & 0.157	  & 0.089	  & 0.082	  & 6.583	  & 2.828	 & 3.035	\\
MCG+07-41-001	 & -0.185	  & 0.058	  & 0.072	  & 32.651	  & 5.932	  & 6.483	  & -0.432	  & 0.043	  & 0.132	  & 36.121	  & 2.281	 & 1.886	\\
MCG-01-25-009	 & -0.174	  & 0.116	  & 0.107	  & 15.698	  & 5.170	  & 6.809	  & 0.041	  & 0.100	  & 0.081	  & 11.856	  & 3.307	 & 4.690	\\
NGC 2841	 & 0.037	  & 0.069	  & 0.067	  & 12.558	  & 1.739	  & 2.606	  & 0.036	  & 0.048	  & 0.050	  & 12.578	  & 1.837	 & 3.073	\\
NGC 2962	 & 0.261	  & 0.093	  & 0.092	  & 5.180	  & 1.993	  & 2.185	  & 0.385	  & 0.066	  & 0.106	  & 4.210	  & 1.572	 & 1.818	\\
NGC 2986	 & 0.052	  & 0.086	  & 0.076	  & 9.652	  & 2.003	  & 2.108	  & 0.014	  & 0.058	  & 0.059	  & 9.667	  & 1.988	 & 1.998	\\
NGC 3608	 & -0.013	  & 0.067	  & 0.069	  & 11.591	  & 2.072	  & 2.745	  & 0.011	  & 0.050	  & 0.052	  & 10.448	  & 1.235	 & 3.472	\\
NGC 4059	 & -0.350	  & 0.158	  & 0.151	  & 7.567	  & 1.664	  & 3.247	  & 0.090	  & 0.108	  & 0.109	  & 5.514	  & 2.517	 & 2.934	\\
NGC 4374	 & 0.102	  & 0.055	  & 0.052	  & 8.667	  & 1.195	  & 1.389	  & 0.081	  & 0.045	  & 0.048	  & 8.851	  & 1.369	 & 1.700	\\
NGC 4472	 & 0.179	  & 0.044	  & 0.045	  & 8.752	  & 1.273	  & 1.466	  & 0.154	  & 0.041	  & 0.038	  & 9.360	  & 1.325	 & 1.346	\\
NGC 4478	 & -0.088	  & 0.079	  & 0.077	  & 9.027	  & 1.641	  & 1.964	  & -0.031	  & 0.051	  & 0.057	  & 8.838	  & 1.882	 & 1.812	\\
NGC 4489	 & 0.428	  & 0.186	  & 0.200	  & 2.539	  & 0.719	  & 0.753	  & 0.167	  & 0.099	  & 0.104	  & 3.350	  & 0.808	 & 2.099	\\
NGC 4493	 & 0.014	  & 0.221	  & 0.224	  & 7.165	  & 3.964	  & 5.585	  & -0.077	  & 0.161	  & 0.160	  & 7.561	  & 4.088	 & 6.693	\\
NGC 4526	 & 0.400	  & 0.061	  & 0.062	  & 1.669	  & 0.215	  & 0.197	  & 0.488	  & 0.036	  & 0.031	  & 1.122	  & 0.279	 & 0.312	\\
NGC 4552	 & -0.040	  & 0.024	  & 0.063	  & 13.481	  & 1.762	  & 2.532	  & 0.056	  & 0.047	  & 0.041	  & 11.834	  & 1.136	 & 2.511	\\
NGC 4649	 & -0.076	  & 0.051	  & 0.055	  & 21.565	  & 2.507	  & 2.533	  & 0.089	  & 0.047	  & 0.048	  & 16.931	  & 2.295	 & 2.219	\\
NGC 4786	 & 0.000	  & 0.099	  & 0.097	  & 7.100	  & 1.684	  & 1.740	  & 0.092	  & 0.077	  & 0.067	  & 6.215	  & 2.690	 & 2.219	\\
NGC 5005	 & \nodata   & \nodata  & \nodata  & \nodata  & \nodata  & \nodata  & \nodata  & \nodata  & \nodata  & \nodata  & \nodata & \nodata	\\
NGC 5061	 & 0.359	  & 0.035	  & 0.056	  & 1.923	  & 0.185	  & 0.054	  & 0.427	  & 0.026	  & 0.026	  & 1.429	  & 0.271	 & 0.202	\\
NGC 5308	 & -0.177	  & 0.063	  & 0.063	  & 11.087	  & 1.923	  & 1.809	  & 0.025	  & 0.054	  & 0.053	  & 9.578	  & 1.758	 & 1.934	\\
NGC 5440	 & 0.057	  & 0.107	  & 0.102	  & 7.415	  & 1.957	  & 2.536	  & 0.248	  & 0.076	  & 0.080	  & 5.453	  & 2.151	 & 2.645	\\
NGC 5490	 & 0.012	  & 0.124	  & 0.120	  & 9.840	  & 2.867	  & 3.296	  & 0.155	  & 0.099	  & 0.097	  & 8.113	  & 3.096	 & 3.338	\\
NGC 5638	 & -0.105	  & 0.079	  & 0.081	  & 13.743	  & 3.146	  & 3.677	  & 0.175	  & 0.054	  & 0.060	  & 9.903	  & 2.322	 & 2.165	\\
NGC 5813	 & 0.027	  & 0.086	  & 0.074	  & 8.692	  & 1.891	  & 1.996	  & 0.369	  & 0.068	  & 0.050	  & 4.921	  & 0.737	 & 2.135	\\
NGC 5846	 & -0.147	  & 0.099	  & 0.047	  & 14.421	  & 2.525	  & 4.724	  & -0.038	  & 0.045	  & 0.053	  & 13.536	  & 2.046	 & 3.490	\\
NGC 6038	 & -0.302	  & 0.194	  & 0.191	  & 12.286	  & 5.161	  & 10.610	  & -0.482	  & 0.186	  & 0.167	  & 13.933	  & 6.198	 & 8.528	\\
NGC 6127	 & -0.178	  & 0.121	  & 0.066	  & 14.917	  & 3.711	  & 6.099	  & -0.122	  & 0.097	  & 0.050	  & 13.975	  & 3.287	 & 5.596	\\
NGC 6411	 & -0.108	  & 0.092	  & 0.091	  & 14.833	  & 3.788	  & 5.096	  & -0.442	  & 0.067	  & 0.067	  & 19.433	  & 4.681	 & 5.155	\\
NGC 6702	 & 0.442	  & 0.121	  & 0.126	  & 1.965	  & 0.379	  & 0.530	  & 0.157	  & 0.082	  & 0.068	  & 2.845	  & 0.478	 & 0.548	\\
NGC 6703	 & 0.078	  & 0.069	  & 0.063	  & 8.186	  & 1.451	  & 1.695	  & 0.213	  & 0.052	  & 0.052	  & 6.829	  & 1.805	 & 1.727	\\
NGC 6986	 & 0.602	  & 0.264	  & 0.320	  & 1.289	  & 0.111	  & 0.067	  & 0.744	  & 0.225	  & 0.122	  & 1.020	  & 2.512	 & 0.179	\\
NGC 7131	 & -0.281	  & 0.134	  & 0.130	  & 23.653	  & 7.315	  & 8.609	  & -0.327	  & 0.112	  & 0.107	  & 24.228	  & 7.778	 & 9.103	\\
NGC 7761	 & 0.164	  & 0.186	  & 0.192	  & 2.907	  & 0.613	  & 0.803	  & 0.268	  & 0.128	  & 0.085	  & 2.670	  & 0.652	 & 0.864	\\
UGC 11149 & \nodata  & \nodata	  & \nodata  & \nodata  & \nodata  & \nodata	  & -0.480	  & 0.251	  & 0.245	  & 9.768	  & 5.339	 & 12.562	\\
UGC 11198	 & -0.238	  & 0.108	  & 0.112	  & 18.449	  & 6.004	  & 5.519	  & -0.204	  & 0.085	  & 0.083	  & 17.861	  & 5.797	 & 5.749	\\
UGC 03787	 & 0.446	  & 0.291	  & 0.419	  & 1.284	  & 0.076	  & 0.082	  & 0.422	  & 0.186	  & 1.729	  & 1.172	  & 0.235	 & 0.190	\\
UGC 04322	 & -0.155	  & 0.155	  & 0.137	  & 11.844	  & 3.818	  & 6.540	  & -0.116	  & 0.108	  & 0.102	  & 11.527	  & 3.927	 & 6.212	\\
\enddata
\end{deluxetable}
\clearpage

\end{document}